\documentclass[useAMS,usenatbib]{mn2e}

\usepackage{graphicx}
\usepackage{amssymb}
\usepackage{amsmath}

\usepackage{natbib}
\usepackage[backref,breaklinks,colorlinks,citecolor=blue]{hyperref}
\usepackage{subfigure}
\usepackage[flushleft]{threeparttable}
\usepackage{etoolbox}

\usepackage{color}

\newcommand{\Ms}{~{\rm M}_\odot}

\newcommand{\bh}{{\rm SMBH}}

\newcommand{\HiGPUs}{\texttt{HiGPUs}}

\makeatletter

\patchcmd{\NAT@citex}
  {\@citea\NAT@hyper@{%
     \NAT@nmfmt{\NAT@nm}%
     \hyper@natlinkbreak{\NAT@aysep\NAT@spacechar}{\@citeb\@extra@b@citeb}%
     \NAT@date}}
  {\@citea\NAT@nmfmt{\NAT@nm}%
   \NAT@aysep\NAT@spacechar\NAT@hyper@{\NAT@date}}{}{}

\patchcmd{\NAT@citex}
  {\@citea\NAT@hyper@{%
     \NAT@nmfmt{\NAT@nm}%
     \hyper@natlinkbreak{\NAT@spacechar\NAT@@open\if*#1*\else#1\NAT@spacechar\fi}%
       {\@citeb\@extra@b@citeb}%
     \NAT@date}}
  {\@citea\NAT@nmfmt{\NAT@nm}%
   \NAT@spacechar\NAT@@open\if*#1*\else#1\NAT@spacechar\fi\NAT@hyper@{\NAT@date}}
  {}{}

\makeatother

\title[The MEGaN project I]{The MEGaN project I. Missing formation of massive nuclear clusters and tidal disruption events by star clusters - massive black hole interactions}

\author[Arca-Sedda M. and Capuzzo-Dolcetta R.]{
M. ~Arca-Sedda$^{1,2}$\thanks{E-mail: m.arcasedda@ari.uni-heidelberg.de}
, R. ~Capuzzo-Dolcetta$^{1}$ 
\\
$^{1}$Dept. of Physics, Sapienza, University of Rome, Piazzale Aldo Moro 5, I-00185, Rome (Italy)\\
$^{2}$Zentrum f\"{u}r Astronomie der Universit\"{a}t Heidelberg, Astronomisches Rechen-Institut, M\"{o}nchhofstr. 12-14, D-69120, Heidelberg (Germany)\\
}

\begin{document}

\date{Revised to }

\pagerange{\pageref{firstpage}--\pageref{lastpage}} \pubyear{2015}

\maketitle

\label{firstpage}

\maketitle

\begin{abstract}
We investigated the evolution of a massive galactic nucleus hosting a super-massive black hole (SMBH) with mass $M_\bh=10^8\Ms$ surrounded by a population of 42 heavy star clusters (GCs). Using direct $N$-body modelling, we show here that the assembly of an NSC through GCs orbital decay and merger is efficiently inhibited by the tidal forces exerted from the SMBH.
The GCs mass loss induced by tidal forces causes a significant modification of their mass function, leading to a population of low-mass ($<10^4$) clusters.
Nonetheless, the GCs debris accumulated around the SMBH give rise to well-defined kinematical and morphological properties, leading to the formation of a disk-like structure. Interestingly, the disk is similar to the one observed in the M31 galaxy nucleus, which has properties similar to our numerical model.
The simulation produced a huge amount of data, which we used to investigate whether the GC debris deposited around the SMBH can enhance the rate of tidal disruption events (TDEs) in our  galaxy inner density distribution. Our results suggest that the GCs disruption shapes the SMBH neighbourhoods leading to a TDE rate of $\sim 2 \times 10^{-4} $yr$^{-1}$, a value slightly larger than what expected in previous theoretical modelling of galaxies with similar density profiles and central SMBHs. The simulation presented here is the first of its kind, representing a massive galactic nucleus and its star cluster population on scales $\sim 100$ pc.
\end{abstract}


\section{Introduction}

In the last twenty years, the Hubble Space Telescope  discovered the existence of very dense and bright nuclei in the centre of many galaxies, called nuclear star clusters (NSCs) or ``resolved stellar nuclei'', with masses up to $10^8\Ms$ 
\citep{boker02,cote06,graham09}. 

NSCs are found in galaxies of all the Hubble types \citep{boker04,Turetal12,georgiev14,denB}, and are characterised by a complex star formation history \citep{walcher06,rossa}.

Very often, galactic nuclei contain at their centre super-massive black holes (SMBHs), with masses in the range $10^6-10^{10}\Ms$ \citep{UrPa,Graham11,VdeBo12,Shankar09,Kormendy13,Merri13,Ems13}. 

While the nuclei of fainter galaxies, with masses below or around $10^{10}\Ms$, seem to be dominated by the presence of a NSC, heavier ones, with masses above $10^{11}\Ms$ seem to contain only SMBHs. Within these limiting values, instead, SMBHs and NSCs co-exist \citep{seth08,LGH,scot}, thus suggesting that NSC-dominated and SMBH-dominated galaxies constitutes a continuous sequence \citep{bekki10}. Due to this, NSCs and SMBHs are often referred to as central massive objects (CMOs).

The existence of scaling relations between SMBHs, NSCs and their host galaxies can give clues about their formation history and evolution. A widely studied relation connects the host galaxy velocity dispersion ($\sigma_g$) and the CMO mass, the so-called $M-\sigma_g$ relation. 

\cite{frrs} supported an SMHBs $M-\sigma_g$ relation  characterised by a steep slope ($M \propto \sigma_g^\alpha; ~\alpha \simeq 4$) whereas for NSCs it is quite shallower, with $\alpha \simeq 2$ \citep{LGH,ERWGD,graham12,georgiev16,tostaCD16}. This would suggest that the two classes of massive objects follow two different evolutionary pathways.

Actually, regarding NSCs, there are two main possible, and debated, formation mechanisms: i) the ``in-situ'' scenario, in which a NSC is thought to form through a series of episodic gas infalls that could drive the formation of a SMBH \citep{davies11}, or accretes onto it \citep{King03,Mil04,King05,McLgh,bekki07,nayakshin,hopkins10,Antonini15,aharon15}; ii) the ``dry-merger'' scenario, in which massive globular clusters (GCs) sink toward the galactic centre due to the action of dynamical friction (df) and merge, raising the formation of a NSC \citep{Trem76, Dolc93, DoMioA, antonini13, AMB, ASCD14a}.

Using  theoretical and statistical approaches, many authors have shown that the dry-merger scenario allows to draw quantitatively scaling relations among the NSCs properties and those of their host galaxy that well fit the observational correlations \citep{antonini13,ASCD14b,gnedin14}. 

Moreover, the inspiralling and merger of GCs in a galaxy has found a significant support by very recent finding of RR Lyrae stars in the centre of our Galaxy \citep{min16}

Using observational data of the dwarf starburst galaxy Henize 2-10, which hosts an SMBH surrounded by 11 massive clusters \citep{reines12,ngu14}, \cite{ASCD15He} have shown that the formation of a NSC is regulated by the gravitational galactic field and the SMBH, and it can occur on relatively short time-scales, $<1$ Gyr. Since the Henize 2-10 clusters are young, with ages $\sim 5$ Myr \citep{chandar03}, these results suggest that NSCs can form on time scales compatible with the SMBH growth.

The dry-merger scenario seems to explain the absence of nucleated regions in dwarf spheroidal galaxies (dSph), where SMBHs are absent, and can provide clues on their DM content, as recently shown by \citep{ASCD16a, ASCD16b}.

As pointed out above, NSCs seem to disappear when the host galaxy mass, $M_g$, exceeds $\sim 10^{11}\Ms$ \citep{frrs}. Moreover, the transition between NSC- and SMBH- dominated galaxies occurs when their masses equal \citep{Neum,ASCD14b}. Therefore, it is possible to determine a mass treshold for the central SMBH above which NSCs cannot form , which is $M_{\rm SMBH} \gtrsim 10^8 \Ms$ \citep{antonini13,ASCD14b}.

Recently, \cite{ASCDS16} have shown that the tidal torques exerted by the SMBH can quench the formation of a NSC if the time-scale over which the SMBH grows is significantly shorter than the GC (formation + df) time-scale.
Let us note that multiple GCs scattering over the SMBH could produce several detectable phenomena, such as the ejection of high and hyper-velocity stars (HVSs) \citep{ASCDS16,fragione15,fragione17}, or tidal disruption events (TDEs).

As opposed to NSC dry-merger scenario, some authors proposed that the absence of nucleated regions in heavy galactic nuclei is due to the {\textit scouring} of the nucleus operated by SMBH binary systems (BHBs). Indeed, during multiple galaxy mergers the SMBHs hosted in the galaxies nuclei are brought together, and their mutual interaction excavates the merger product nucleus. This causes a mass deficit in the innermost galactic region, which depends on the SMBH final mass and the number of merging experienced by the host galaxy \citep{merritt06}. 
Moreover, stalled satellites can further enlarge the size of the galaxy core, although the origin of such satellite can be related to multiple SMBHs, debris of a minor merger event or massive star clusters \citep{bekki10,bonfini16,dosopoulou17,donnari17}. 

However, it is worth noting that these two mechanisms do not necessarily act in competition, as they operate on two different time-scales. Indeed, if the post-merged galaxy contains a population of massive star clusters, after the BHB merger they can undergo df and accumulate into the galactic centre, contributing to the NSC assembly as occur in lighter galaxies. On the other hand, if the final SMBH and the galaxy are sufficiently massive and dense, their tidal forces can disrupt the infalling clusters and quench the NSC formation. 
In this regard, E+A galaxies are particularly interesting objects. Indeed, these E/S0 galaxies are thought to underwent a starburst $\sim 1$ Gyr triggered by a merger event  \citep{dressler83,zabludoff96,quintero04}. Recent observations suggest that some of these galaxies host a population of young massive clusters formed during and immediately after the post-starburst phase \citep{yang04}. Therefore, E+A galaxies seem to be the perfect example of post-merged systems in which star clusters orbital evolution may contribute, or not, to the formation of a bright nucleus.

In this context, we present a direct $N$-body modelling of a massive galactic nucleus ($M_g=10^{11}\Ms$) containing 42 GCs with masses in the range $3\times 10^5-2\times 10^6\Ms$ and an SMBH of mass $10^8\Ms$. The results of the simulations allowed to draw the GCs orbital evolution and the SMBH role at unprecedented level of detail, thus representing reliably the interplay between the SMBH and its neighbourhood. 

In particular, we focused our attention on several key aspects of galactic nuclei evolution:
\begin{itemize}
\item possible formation of a massive NSC;
\item production of high (hyper)-velocity stars during the GCs-SMBH gravitational collisions;
\item probability of having TDEs after GCs-SMBH interactions;
\item ejection of black hole binaries (BHBs) from the GCs cores during their infall and possible decrease of their coalescence time-scale;
\item interaction between the central SMBH and intermediate mass black holes (IMBHs) transported within the infalling GCs;
\item implications for extreme mass-ratio inspirals (EMRIs);
\end{itemize}

This paper is intended as the first of a series of three which would deepen the points indicated above. This paper I aims at investigating the direct consequence of the GC orbital evolution in the combined galaxy+SMBH field to the spatial and kinematical structure of the inner galactic part. The other papers will mainly deal with, respectively, the formation of black hole binaries and their eventual coalescence and the possible effects of IMBHs on the SMBH dynamical evolution (paper II) and the topics of hyper velocity stars generation (paper III).

The paper is organised as follows: in Section \ref{model} we introduce the galaxy model and the GCs orbital and structural properties; in Section \ref{results} we introduce the results of the numerical simulation with particular focus on the competing action of dynamical friction and tidal disruption processes; in Section \ref{discussion} we discuss some implications of our results: in particular we show in Section \ref{NSC} that the GC orbital evolution does not drive the formation of an observable NSC when the SMBH is as massive as the one considered here while in Sect. \ref{TDE} we focus the attention on the impact of GC-SMBH interaction in favouring or preventing TDEs; finally, Sect. \ref{conclusions} is devoted to the conclusions.

\section{Model}
\label{model}

In this work we simulated the evolution of 42 GCs moving in the inner region of a galaxy hosting an SMBH with mass $10^8$ M$_\odot$. The simulation has been performed in the framework of the ``\textit{Modelling the Evolution of Galactic Nuclei}'' (MEGaN) project.
 
We model the galaxy and the clusters by particles, finding a good balance between the total computational load and the reliability of the system representation.

To model the galaxy, we used a truncated Dehnen model \citep{Deh93}
\begin{equation}
\rho_{\rm D}(r)=\frac{(3-\gamma)M_g}{4\pi r_g^3}\left(\frac{r}{r_g}\right)^{-\gamma}\left(1+\frac{r}{r_g}\right)^{-4+\gamma},
\label{dens}
\end{equation}
with $M_g$ the total galaxy mass, $r_g$ its scale radius and $\gamma$ tunes the steepness of the profile.
We consider a galaxy with total mass $M_g=10^{11}$ M$_\odot$. According to the $M_g-M_\bh$ relation provided by \cite{scot}, such a value of $M_g$ implies an SMBH mass of $\sim 10^8$ M$_\odot$. In order to generate a reliable model, we chose $r_g=2$ kpc and $\gamma=0.1$, which result into a galaxy effective radius $R_e = 5.3$ kpc. 
As shown by \cite{Merri13}, $\gamma=0.5$ is the maximum value allowed to isotropic distribution functions around a masive black hole.  Therefore, our choice of $\gamma = 0.1$ implies some anisotropy of the stars distribution function  within the SMBH influence radius.

On the other hand, galaxy merger and SMBH pairing and collisions can cause a significant flattening of the merger product density profile, leading to $\gamma$ values smaller than 0.3 \citep{merritt06}.

Sampling this galaxy model by particles would require $\sim 10^{11}$ particles, a number exceedingly large to be simulated even with the most advanced computational devices available on the market.
Due to this, we restricted our galaxy model to the thinner $150$ pc, adopting a modification of the density profile in Eq. \ref{dens}
\begin{equation}
\rho(r)=\frac{\rho_{\rm D}(r)}{{\rm cosh}(r/r_{\rm cut})},
\end{equation}
where ${\rm cosh}(r/r_{\rm cut})$ is the usual hyperbolic cosine function and $r_{\rm cut}=150$ pc. This choice allows us to model the galaxy nucleus with a mass of $2.8\times 10^8$ M$_\odot$.

We selected randomly the 42 GCs masses in the range $0.3-2\times 10^6$ M$_\odot$, obtaining a total mass $5\times 10^7$ M$_\odot$ for our sample of GCs. 

\begin{table*}
\caption{}
\centering{Properties of the GCs sample}
\begin{center}
\begin{tabular}{lcccccccccc}
\hline
\hline
\multicolumn{1}{c}{GC}  & $W_0$ & $R_t$ & $R_c$ & $M_{\rm GC}$ & $r_{\rm GC}$ & $v_{\rm GC}$ & $e$ & $t_{\rm df}$ & $M_{\rm GC,f}/M_{\rm GC}$ & $N_{\rm GC}$\\
\multicolumn{1}{c}{name} &       & (pc)  & (pc)  & $10^6$ M$_\odot$ & (pc) & km s$^{-1}$ &  & (Gyr)& ($\%$) & \\ 
\hline
GC1 &7.54 &10.9 &0.207 &1.05 &71.4 & 121 & 0 & 0.295 & 4.94 & 10445 \\
GC2 &7.13 &17 &0.443 &0.906 &117 & 89.4 & 0.527 & 0.509 & 30.8 & 8995 \\
GC3 &7.66 &23.8 &0.424 &1.68 &134 & 87.7 & 0.483 & 0.446 & 46.8 & 16671 \\
GC4 &7.08 &13.7 &0.378 &1.89 &74.3 & 115 & 0.974 & 0.0755 & 2.61 & 18754 \\
GC5 &7.89 &16.5 &0.257 &1.1 &107 & 91.6 & 0.553 & 0.368 & 60.5 & 10966 \\
GC6 &7.28 &15.1 &0.35 &0.452 &132 & 24.5 & 0.884 & 0.633 & 1.03 & 4492 \\
GC7 &7.71 &23.2 &0.397 &1.69 &130 & 93.5 & 0.687 & 0.336 & 6.18 & 16792 \\
GC8 &7.78 &20.8 &0.345 &1.07 &136 & 78.2 & 0.177 & 0.804 & 3.95 & 10634 \\
GC9 &6.88 &15.2 &0.477 &1.87 &82.6 & 91.3 & 0.32 & 0.204 & 2.67 & 18554 \\
GC10 &6.58 &22 &0.812 &1.16 &140 & 33.6 & 0.785 & 0.432 & 2.35 & 11560 \\
GC11 &7.33 &22.6 &0.502 &1.25 &140 & 86.2 & 0.411 & 0.629 & 3.41 & 12412 \\
GC12 &7.49 &16.2 &0.32 &1.62 &92.6 & 93 & 0.478 & 0.239 & 22.6 & 16054 \\
GC13 &7.27 &14.9 &0.347 &0.85 &105 & 78 & 0.129 & 0.618 & 41 & 8446 \\
GC14 &6.54 &15.4 &0.58 &0.719 &115 & 76.9 & 0.125 & 0.806 & 40.1 & 7137 \\
GC15 &6.76 &25.1 &0.839 &1.66 &142 & 61.4 & 0.29 & 0.586 & 3.2 & 16530 \\
GC16 &7.74 &12.7 &0.214 &1.28 &78.4 & 121 & 0 & 0.305 & 3.84 & 12716 \\
GC17 &7.01 &8.78 &0.257 &1.86 &47.7 & 127 & 0.729 & 0.051 & 22.9 & 18517 \\
GC18 &6.5 &5.01 &0.194 &0.583 &40.1 & 131 & 0.561 & 0.0997 & 6.73 & 5791 \\
GC19 &6.45 &16.6 &0.665 &0.834 &118 & 69.8 & 0.0695 & 0.801 & 22.8 & 8284 \\
GC20 &6.11 &14.4 &0.727 &0.33 &139 & 85 & 0.38 & 1.56 & 48.9 & 3273 \\
GC21 &7.45 &13.9 &0.283 &1.49 &81.5 & 90.9 & 0.31 & 0.234 & 15.9 & 14840 \\
GC22 &7.1 &19.7 &0.53 &1.38 &119 & 72.6 & 0.00601 & 0.599 & 8.1 & 13721 \\
GC23 &7.14 &19.1 &0.497 &1.7 &107 & 93.5 & 0.616 & 0.26 & 0.626 & 16836 \\
GC24 &6.96 &11.6 &0.352 &1.3 &71.5 & 122 & 0 & 0.257 & 12.9 & 12871 \\
GC25 &6.63 &16 &0.574 &0.613 &126 & 83.1 & 0.32 & 0.903 & 15.7 & 6090 \\
GC26 &6.66 &26.2 &0.921 &1.97 &140 & 55.8 & 0.407 & 0.462 & 28.4 & 19521 \\
GC27 &7.78 &19.6 &0.324 &1.24 &122 & 36.7 & 0.743 & 0.343 & 3.92 & 12353 \\
GC28 &7.88 &8.85 &0.138 &1.27 &54.7 & 136 & 0 & 0.163 & 18 & 12598 \\
GC29 &6.02 &26.8 &1.46 &1.79 &148 & 93.3 & 0.641 & 0.426 & 31.2 & 17756 \\
GC30 &7.2 &18.5 &0.456 &1.03 &122 & 93.9 & 0.686 & 0.421 & 1.03 & 10226 \\
GC31 &6.43 &14.4 &0.583 &0.612 &114 & 56.6 & 0.392 & 0.711 & 4.6 & 6073 \\
GC32 &7.57 &15.1 &0.284 &0.845 &107 & 77.5 & 0.111 & 0.647 & 14.8 & 8397 \\
GC33 &6.71 &9.43 &0.323 &1.03 &62.5 & 135 & 0 & 0.236 & 1.52 & 10258 \\
GC34 &7.17 &19.1 &0.484 &1.08 &124 & 100 & 0.92 & 0.298 & 5.14 & 10776 \\
GC35 &7.25 &11.5 &0.272 &0.486 &98 & 62.1 & 0.316 & 0.683 & 8.28 & 4825 \\
GC36 &7.09 &19.7 &0.536 &1.25 &122 & 34 & 0.778 & 0.328 & 8.1 & 12369 \\
GC37 &6.47 &24.3 &0.957 &1.69 &137 & 79.6 & 0.217 & 0.576 & 4.11 & 16787 \\
GC38 &7.6 &14.5 &0.268 &0.334 &140 & 85.2 & 0.379 & 1.56 & 8.97 & 3318 \\
GC39 &6.16 &23.9 &1.17 &1.41 &143 & 84.3 & 0.336 & 0.636 & 2.45 & 13990 \\
GC40 &7.18 &23 &0.575 &1.56 &133 & 87.6 & 0.485 & 0.458 & 13.2 & 15468 \\
GC41 &7.74 &23.7 &0.401 &1.39 &142 & 71.1 & 0.0492 & 0.8 & 31.7 & 13768 \\
GC42 &7.2 &4.55 &0.112 &0.478 &38.9 & 141 & 0.781 & 0.0828 & 8.1 & 4747 \\
\hline
\end{tabular}
\end{center}
\begin{tablenotes}
\item Column 1: GCs name. Column 2: value of the adimensional potential well. Column 3: GC tidal radius. Column 4: GC core radius. Columns 5-7: GC mass, initial position and velocity. Column 7: GC orbital eccentricity. Column 8: dynamical friction timescale according to Eq. \ref{tdf}. Column 9: GC mass percentage by the end of the simulation. Column 10: number of particles used to model the GCs.
\end{tablenotes}
\label{GCS}
\end{table*}

To model the whole system (galaxy+GCs) we used a total number of particles equal to $2^{20}$. Moreover, we assumed for particles in the galaxy model an individual mass 5 times larger than for particles in the GCs. This choice allowed us to model the smallest cluster with more than 2000 particles, thus ensuring an evaporation time-scale, which is the time-scale over which two-body relaxation drives the GC disruption, $\simeq 3$ Gyr, sufficiently longer than the simulated integration time.
We performed several test runs at varying the ratio of galaxy to GC particle masses, finding that the choice of a ratio equal to 5 gives a very reliable simulation outcome. 

The GCs initial positions and velocities have been selected self-consistently, according to the background density distribution.

Once the GCs positions have been assigned, it is possible to estimate their tidal radius, $R_t$, which defines the region within stars are bounded to the GC.
A general way to estimate such length-scale is the following
\begin{equation}
R_t = \left(\frac{GM}{\omega^2+{\rm d}^2U/{\rm d}r^2} \right)^{1/3}
\label{rtid}
\end{equation}

Each GC has been modelled via a King density profile \citep{King}, which is defined by the adimensional potential well $W_0$, the GC core radius $R_c$ and its total mass $M$. 
In order to provide reliable models, we selected $W_0$ in the range $6-8$, since only GCs with a sufficiently deep potential well can arrive near the central SMBH without being disrupted by the strong tidal forces.

\begin{figure}
\centering
\includegraphics[width=8cm]{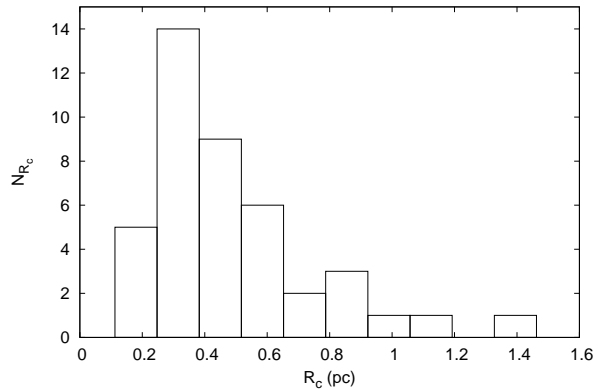}
\caption{GCs core radius initial distribution.}
\label{Rcdist}
\end{figure}

The knowledge of $W_0$ and $R_t$ allows to estimate the core radius, $R_c$, thanks to the correlation between $W_0$ and the $c$ concentration parameter, defined as $c=R_t/R_c$.
Figure \ref{Rcdist} shows the number distribution of the core radii evaluated this way.

Table \ref{GCS} summarizes the main parameters of our GC models. 

Our simulation has been run on two different composite platforms: i) ASTROC9, a desktop computer hosting 2 Xeon X5650 processor and 4 RADEON HD 7990 graphic processing units (GPUs); ii) ASTROC16a, hosting  2 Xeon E5-2623v3 and 4 NVIDIA Titan X GPUs. We used the direct $N$-body code \HiGPUs \hspace{0.2cm}
\citep{Spera}, a highly parallel, direct summation, 6$^{th}$ order, Hermite integrator that  implements block-time steps. 

\section{Results}
\label{results}

In order to follow the GC evolution it is important to determine three important quantities: the GC centre of mass (COM), its tidal radius and its bounded mass. While this is simple for any spherically symmetric system, where the COM coincides with the centre of density (COD), for systems suffering intense tidal force this evaluation becomes much more complicated.
In our case, we developed an algorithm that using the COM as starting point, evaluates in a recursive way the centre of density. As shown in \cite{ASCDS16} (see their Fig. 1), our approach works very well even in the case of a severely warped GC. 
Once the infalling GC COD is evaluated correctly, another important quantity is its tidal radius $R_t$. According to Eq. \ref{rtid}, its evaluation depends on the GC bounded mass, $M$, that in turn depends on  $R_t$. 

Therefore, to give a proper estimate of the tidal radius, we firstly use the total GC mass in Eq. \ref{rtid}, then we evaluate the GC mass enclosed within $R_t$ and use it to re-estimate $M$. After naming $M_0$ the first guess for the total mass of the GC, we followed the scheme

\begin{align}
M_0 \rightarrow R_{tid,0}(M_0) \rightarrow M_1(R_{tid,0}) \rightarrow R_{tid,1} \rightarrow ... \\ \nonumber
\rightarrow M_{i-1}(R_{tid,i-2}) \rightarrow R_{tid,i-1}(M_{i-1})\rightarrow M_i(R_{i-1});
\end{align}

which we stop when the relative variation of the GC mass falls below $0.001$.

\subsection{Dynamical friction and tidal disruption}
\label{dftd}

During their motion, GCs undergo dynamical friction (df), which causes their orbital decay toward the centre of the galaxy \citep{Cha43I,Trem76,Dolc93}. 

The failure of the classical treatment developed by \cite{Cha43I} in describing the df in a dense galactic nucleus moved several authors to develop semi-analytical treatments, that have been robustly tested against numerical experiments and reproduce satisfactorily the evolution of massive satellites spiralling around dense galactic nuclei and SMBHs \citep{AntMer12,ASCD14a,petts16,dosopoulou17}.
The time-scale over which this process occur is well described by the following formula, deeply discussed in \cite{ASCD14a} and \cite{ASCD15He}

\begin{equation}
t_{\rm df} ({\rm Myr})= t_0g(e,\gamma)\left(\frac{M_{\rm GC}}{M_g}\right)^{-0.67}\left(\frac{r_{\rm GC}}{r_g}\right)^{1.76},
\label{tdf}
\end{equation}

where $M_{\rm GC}$ is the cluster mass, $r_{\rm GC}$ its radial position within the host (spherical) galaxy and $t_0$ is  given by

\begin{equation}
t_0 ({\rm Myr}) = 0.3 \left(\frac{r_g}{1{\rm kpc}}\right)^{3/2}\left(\frac{10^{11}{\rm M}_\odot}{M_g}\right)^{1/2}.
\end{equation}

The function $g(e,\gamma)$ is given by

\begin{equation}
g(e,\gamma)=(2-\gamma)\left[a_1\left(\frac{1}{(2-\gamma)^{a_2}}+a_3\right)(1-e) + e\right],
\end{equation}
with $a_1= 2.63 \pm 0.17$, $a_2 = 2.26 \pm 0.08$ and $a_3=0.9 \pm 0.1$.

\begin{figure}
\centering 
\includegraphics[width=8cm]{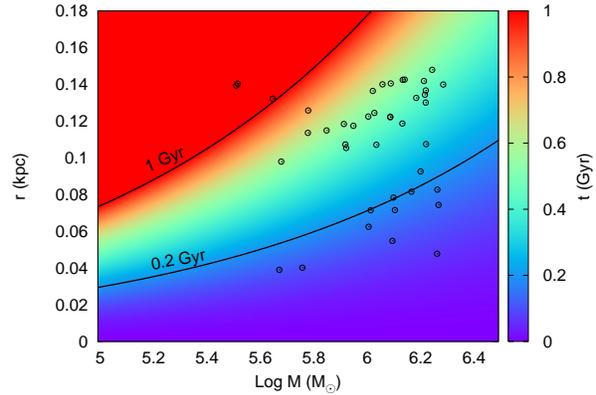}
\caption{colour map of the $t_{\rm df}$ at varying $M_{\rm GC}$ and $r_{\rm GC}$. Open black circles represent our modelled GCs. The map is obtained assuming an initial eccentricity $e=0.5$. An estimate of the GC $t_{\rm df}$ is also listed in Table \ref{GCS}.}
\label{mapDF}
\end{figure}

Figure \ref{mapDF} shows a surface map that describes how the df time-scale varies at varying the GC mass and initial position, according to Eq. \ref{tdf}. Our GC sample is also represented. 

It is worth noting that more than $25\%$ of all the GCs
have dynamical friction time-scales smaller than $200$ Myr. 

On the other hand, as the clusters travel within the host galaxy, the tidal torques induced by the SMBH and the galactic background induce a shattering of the incoming GCs and, in some cases, disrupt them before they approach the SMBH. We found that $57\%$ of the GCs lost more than $90\%$ of their initial mass after $223$ Myr, making clear the role played by tidal heating in determining the mass deposited around the SMBH. We stopped our simulation after $\simeq 290$ Myr, when the intense action of tidal forces have almost completely destroyed the GCs in our sample.

The mass loss caused by tidal torques can be monitored assuming that the GC is at any time described by a King density profile. 

Assuming a King profile, the GC mass satisfies the relation \citep{king62}

\begin{equation}
M = \frac{R_t\sigma^2}{2G},
\label{msigma}
\end{equation}

where $\sigma$ is the GC 1D velocity dispersion and $R_t$ its tidal radius. Coupling Eqs. \ref{rtid} and \ref{msigma} we get

\begin{equation}
M = \frac{\sigma^3}{2\sqrt{2}G}\left(\omega^2 + \frac{{\rm d}^2U}{{\rm d}r^2}\right)^{-1/2},
\label{mtid}
\end{equation}

According to our galaxy model, the gravitational potential generated by the background galaxy and the SMBH is given by

\begin{equation}
U = \frac{GM_{\rm SMBH}}{r}+\frac{GM_g}{(2-\gamma)r_g}\left[1-\left(\frac{r}{r+r_g}\right)^{2-\gamma}\right].
\end{equation}

Therefore, for a Dehnen model we have

\begin{equation}
\omega^2 = \frac{GM_{\rm SMBH}}{r^3}+\frac{GM_g}{r^3}\left(\frac{r}{r+r_g}\right)^{3-\gamma},
\end{equation}
and
\begin{equation}
\frac{{\rm d}^2U}{{\rm d}r^2} = \frac{2GM_{\rm SMBH}}{r^3}-\frac{GM_g}{r^\gamma\left(r+r_g\right)^{4-\gamma}}\left[(1-\gamma)r_g-2r\right].
\end{equation}

Equation \ref{tdf} can be used to describe how the GC radial position $r$ evolves as a function of the time. In particular, a GC that moves from an initial radial position $r_0$ to $r$ after a time $t$, is given by the solution of the following equation

\begin{equation}
t_{\rm df}(r_0)-t_{\rm df}(r) = t,
\end{equation} 

which is

\begin{equation}
r = r_0 \left(1-\frac{t}{t_{\rm df}(r_0)} \right)^{0.57}.
\label{rtdf}
\end{equation}

Combining Eqs. \ref{mtid}-\ref{rtdf} allows to obtain the time evolution of the GC mass, $M$. It is easily seen that

\begin{equation}
M = \frac{\sigma^3}{2\sqrt{2}G}\left[\frac{GM_{\rm SMBH}}{r^3}\left(1+\frac{M_g}{M_{\rm SMBH}}\left(\frac{r}{r_g}\right)^{3-\gamma}\right)\right]^{-1/2},
\end{equation}

which implies

\begin{equation}
M \propto \left(1-\frac{t}{t_{\rm df}(r_0)} \right)^{0.9}.
\end{equation}
	
When the galaxy does not contain a central SMBH, instead, the latter equation reduces to
\begin{equation}
M \propto \left(1-\frac{t}{t_{\rm df}(r_0)} \right)^{0.3\gamma}.
\end{equation}

It should be noted that Eq. \ref{mtid} holds only for nearly circular orbits, and under the additional assumption that the value of $\sigma$ remains nearly constant along the trajectory. So, the above relations represent just a rough estimate of how much mass should be dispersed along the GC trajectory, and only detailed numerical modelling allows a detailed description of the effects of tidal forces.

\begin{figure}
\centering
\includegraphics[width=8cm]{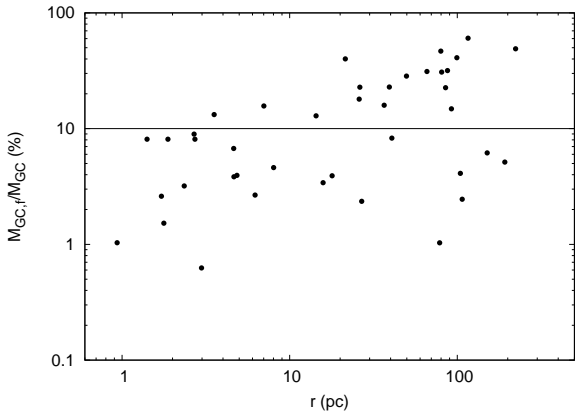}
\caption{Percentage of the GCs masses as a function of their radial distance to the central SMBH at $T =224$ Myr.}
\label{GCdist}
\end{figure}

Due to this, we show in Fig. \ref{GCdist} the percentage of the GCs bounded mass at the end of simulation.

\begin{figure}
\centering
\includegraphics[width=8cm]{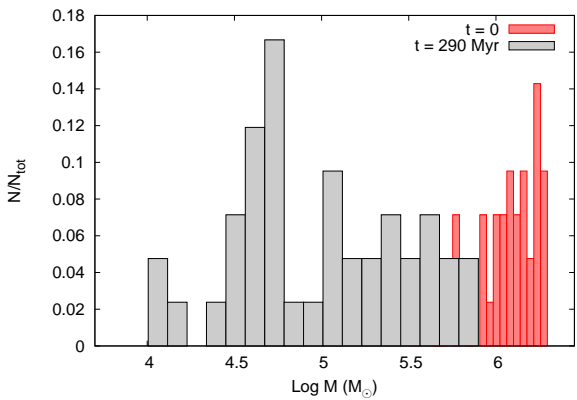}
\caption{Distribution of masses of GCs at the beginning and the end of our simulation.}
\label{remndistri}
\end{figure}
We found that more than $50\%$ of the clusters have masses smaller than one tenth of their initial mass, thus highlighting how much tidal forces affect the growth of the galactic nucleus in this case. 

It is worth noting that mass loss shapes significantly the mass distribution of the GCs. Figure \ref{GCdist} shows the GCs distance to the SMBH as a function of their masses after $220$ Myr. Although the correlation is very weak, it seems that the lighter the cluster the smaller the distance to the SMBH. This behaviour seems to be at odds with the normal expectations of mass segregation, by which the most massive bodies tend to concentrate into the galactic nucleus. Actually, this plot highlights the competitive action of df and tidal heating. Indeed, df drags the most massive clusters toward the SMBH, enhancing, in turn, the tidal torque as the distance decreases. The net result is that we find the lightest clusters nearer to the SMBH while the heavier remain in an outer shell, driving the GCS in a ``\textit{anti-mass segregation}'' state.

The two dominant sources of tidal forces are the central SMBH, and the background galaxy. 
The natural length scale over which the SMBH force dominates over the galactic background is the influence radius, given by

\begin{equation}
r_{\rm inf} = \frac{GM_{\rm SMBH}}{\sigma_g^2},
\end{equation}

with $\sigma_g$ the galaxy central 3D velocity dispersion. In our model $r_{\rm inf} \simeq 13\pm 5$ pc, therefore we expect that GCs suffer tidal forces from the SMBH only when they approach at a distance comparable to $r_{\rm inf}$. Looking at Fig. \ref{GCdist}, it is evident that $16$ out of the $42$ clusters move within $10$ pc, and are likely warped mostly by the SMBH. On the other hand, 9 of the remaining clusters have masses about $10\%$ of their initial values,
thus pointing out the importance of the background galaxy in shaping the structural evolution of the clusters.

The combined tidal action of the SMBH and the galactic background changes significantly the GCs mass distribution. Indeed, by the end of the simulation all the GCs have masses smaller than $6\times 10^5\Ms$ the low-mass cutoff of the distribution shifts to $10^4 \Ms$. Moreover, the distribution of masses above
$3\times 10^4 \Ms$ is well fitted by either an exponential mass function 

\begin{equation}
f(M) = A\exp(-M/B),
\end{equation}

with $A=(16\pm 2)$ M$_{\odot}^{-1}$ and $B=(8.2\pm 1.2) \times 10^4$ M$_\odot$, or by a power-law 
\begin{equation}
f(M) = a\left(\frac{M}{10^5~{\rm M}_\odot}\right)^b,
\end{equation}

with $a = 4.2\pm 2.0$ M$_\odot^{-1}$  and $ b= -0.78\pm 0.14$. 
It is worth noting that after $\sim 220$ Myr only 17 GCs have masses above $10^5$ M$_\odot$. 

This suggests that a massive galaxy likely hosts, in its nucleus, a population of relatively small GCs, characterised by the mass function shown in Fig. \ref{remndistri}.

\section{Discussion}
\label{discussion}

As we showed in the last section, the GC orbital evolution is notably shaped by the presence of the central SMBH, which has a shattering effect on them. 

Several astrophysical processes can be driven by a strong SMBH-GC gravitational collision, such for instance the 
ejection of high-velocity stars, enhancement of stellar disruption by the SMBH. In this paper, using the data provided by our simulation we try to determine some information about these phenomena and their consequences.

\subsection{Is a nuclear star cluster forming around such a massive BH?}
\label{NSC}

The formation of nuclear star clusters by decay and merging of globular clusters has been tested in dwarf \citep{ASCD16a,ASCD16b} and mid-weight galaxies \citep{AMB, mastrobuono14, perets14, ASCD14b, ASCD15He}, whereas a number of recent works have argued that this process works inefficiently in high-mass galaxies \citep{antonini13,ASCD14b,ASCDS16}. 

However, most of the previous works limited their models to about 10 GCs moving around an SMBH, due to the computational load required to simulate a massive galactic nucleus.
In this paper we model the entire galactic nucleus, showing that tidal forces in galaxies hosting a SMBH with mass above $10^8$ M$_\odot$ are sufficiently high to inhibit the formation of a detectable NSC.

As shown in Fig. \ref{GCdist}, a number of GC remnants penetrate the inner region of the galaxy, reaching distances smaller than 10 pc from the SMBH. Therefore, GC debris may, in principle, leave a fingerprint in the SMBH surroundings.

\begin{figure}
\centering
\includegraphics[width = 8cm]{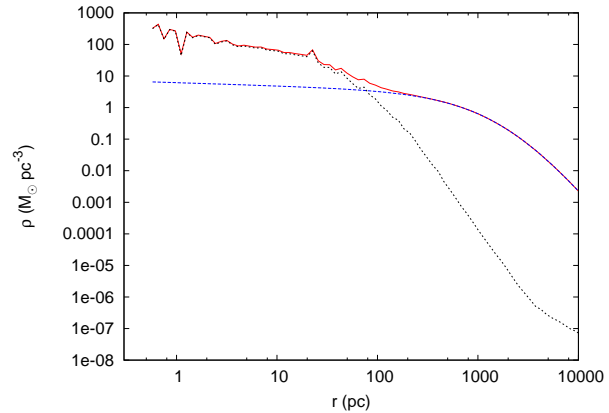}
\caption{The blue curve is the initial galaxy density profile. The dotted curve is the GC density profile at $220$ Myr, and the red curve is the global (galaxy + GC) density profile at $220$ Myr.}
\label{densglob}
\end{figure}

Indeed, the GCs evolution causes a significant flattening of the global three dimensional velocity dispersion profile, which passes from a value, averaged over the inner 20 pc, of $\sim 500$ km s$^{-1}$ to $\sim 100$ km s$^{-1}$ by the end of the simulation. 
Moreover, GC orbital infall and disruption lead to an evident central increase in the spatial density profile, as shown in Fig. \ref{densglob}.

\begin{figure}
\centering
\includegraphics[width=8cm]{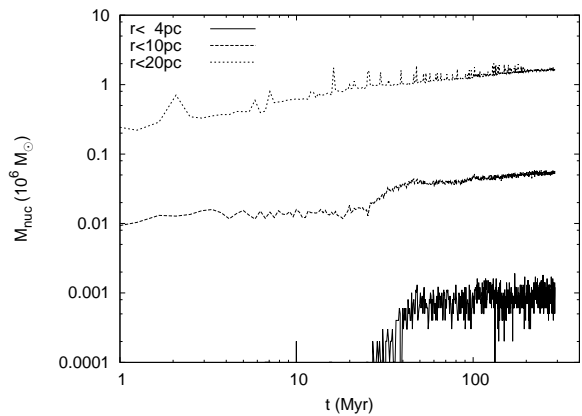}
\caption{Time evolution of the mass deposited around the SMBH at different radii.}
\label{NSCgr}
\end{figure}

A relevant parameter that can be used to determine whether the GCS orbital evolution can give rise to a NSC is the amount of mass deposited around the SMBH. Figure \ref{NSCgr} shows the mass initially bound to the GCs, accumulated at 4, 10 and 20 pc from the SMBH as a function of the time.
It is worth noting that the galaxy mass enclosed within 4 pc according to our galaxy model is $\sim 1500\Ms$, a value compatible with the GCS deposited mass. This would represent a first hint on the weak detectability of a possible NSC.

Observationally, a NSC in a galactic nucleus is identified as an evident edge in the 
host galaxy surface brightness profile \citep{cote06,Turetal12,georgiev14,ASCD15He}.
However, we did not found any evident edge neither in our model surface density profile, nor in the projected radial velocity profile, which is shown in Fig. \ref{beta}.
Therefore, our results suggest that the central SMBH and its surrounding act as a barrier, preventing the NSC formation and leading to an insufficient amount of GCs debris around the SMBH.
Nonetheless, the interactions between the SMBH and the GCs are strong enough to suggest that a number of interesting phenomena can occur, such as BHB coalescence, TDEs and GWs emissions by EMRIs.

\subsubsection{Central structure morphology and kinematics}
 
In this section we investigate the kinematical and morphological properties of the very inner region of the galaxy studied, at distances below 5 pc from the central SMBH.

\begin{figure}
\centering
\includegraphics[width=8cm]{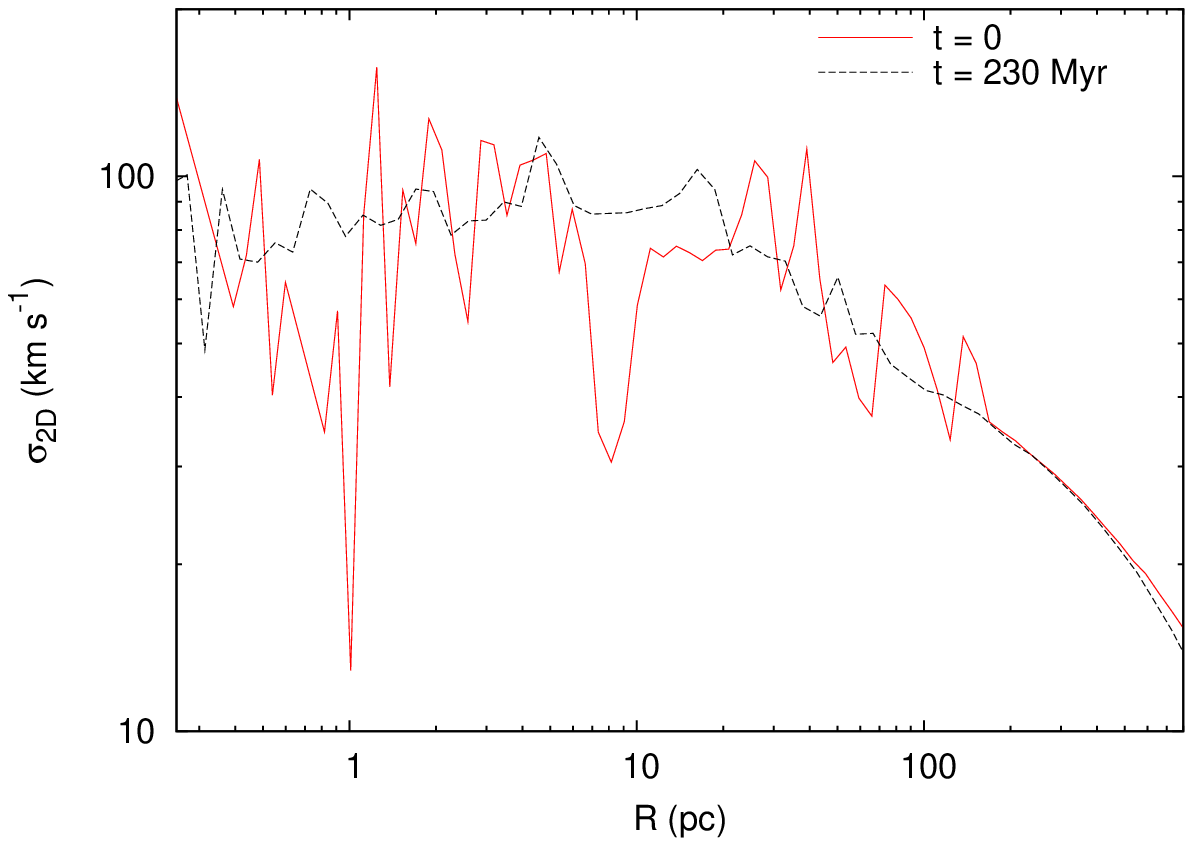}
\includegraphics[width=8cm]{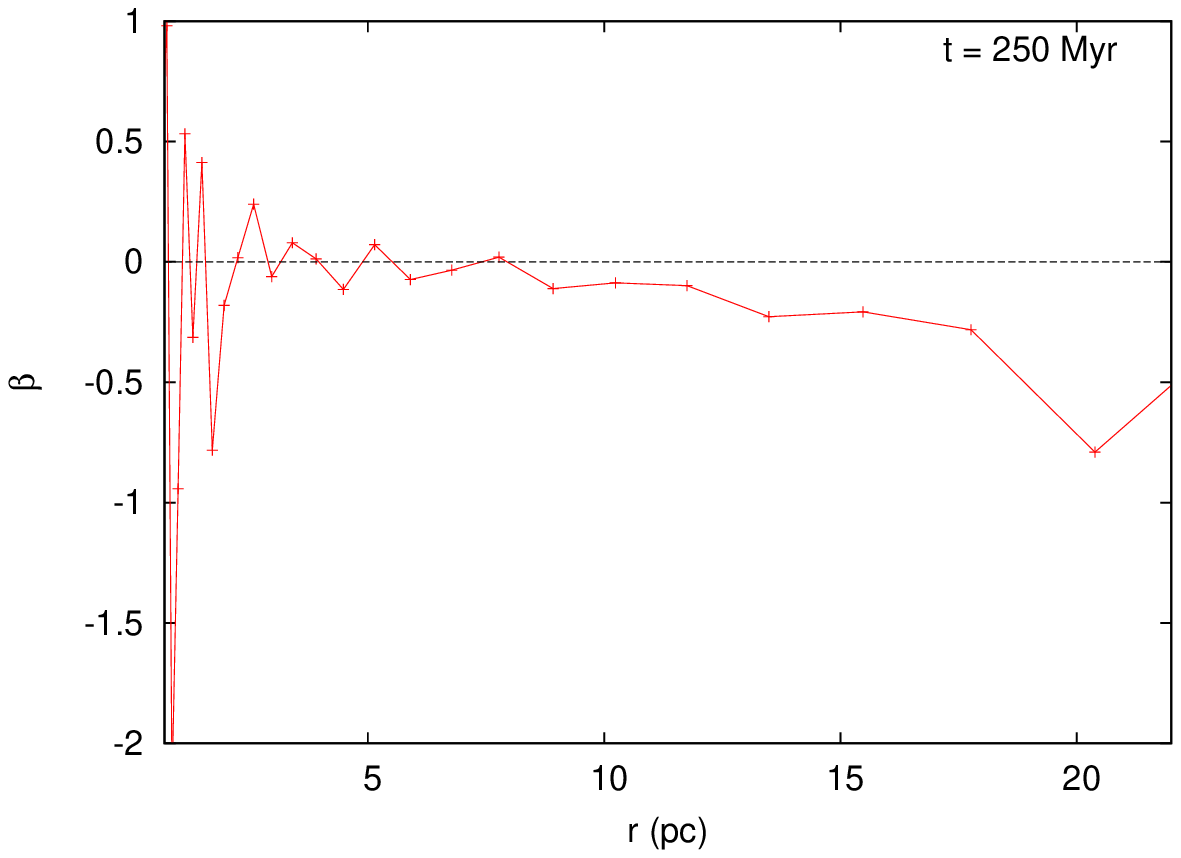}
\includegraphics[width=9cm]{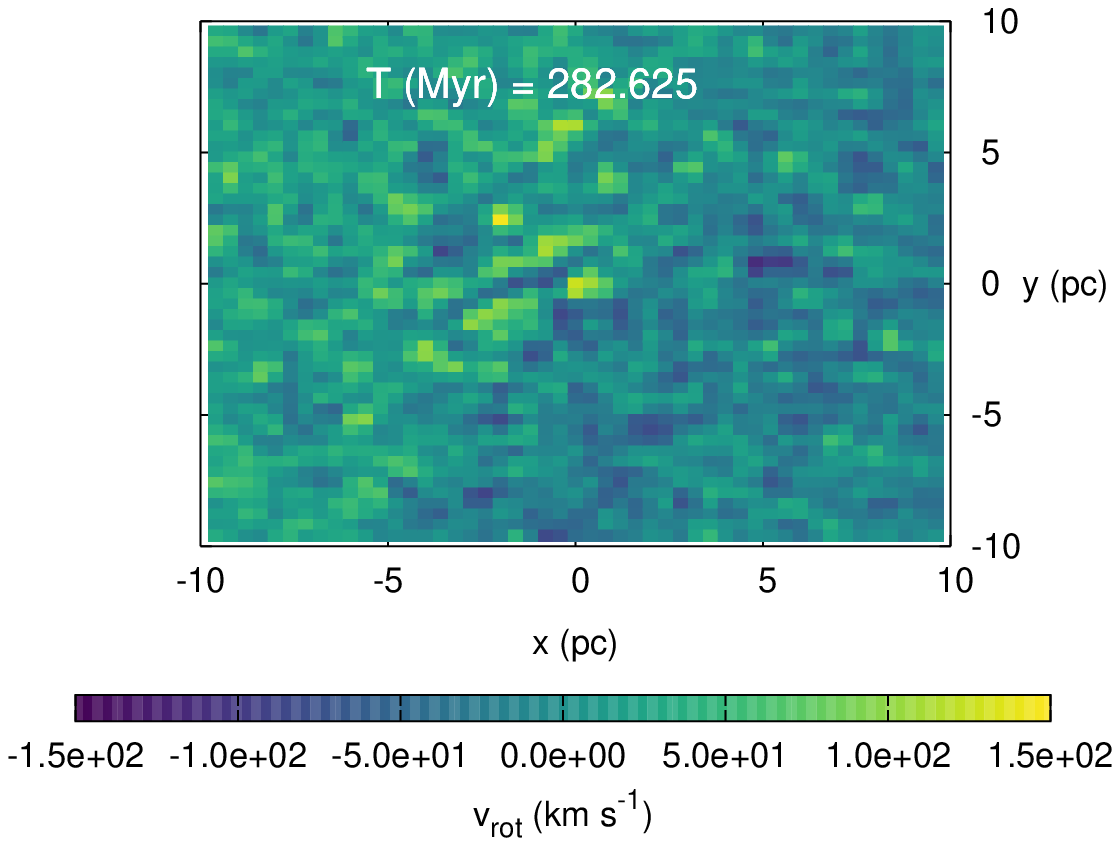}
\caption{Top panel: Surface velocity dispersion of the whole system at $T=0$ (red straight line), and at $T=224$ Myr (black dotted line). Central panel: The value of the anisotropy parameter $\beta$ as a function of galactocentric distance at the time labelled. The dotted black line represent a 0 offset. Bottom panel: Mean velocity, oriented along the z-direction, within the inner $10$ pc around the SMBH. It is evident a mild rotation with amplitude $\sim 100$ km s$^{-1}$.}
\label{beta}
\end{figure}

In central panel of Fig. \ref{beta} we show the time evolution of the $\beta$ anisotropy parameter. This parameter is defined as $\beta=1-(\sigma_t/2\sigma_r)^2$, where $\sigma_t$ and $\sigma_r$ represent the tangential and radial velocity dispersions, respectively. After $\sim 300$ Myr, our galaxy+GCs model is characterised by $\beta \simeq 0$ within $5$ pc from the SMBH, which implies an almost isotropic configuration, while it declines toward negative values outward, showing a predominance of tangential motion at the edge of the galactic nucleus.

Another important set of parameters that can be used to constrain the galaxy morphology is that of the three principal moments of inertia, $I_1>I_2>I_3$, which allow to discriminate between spherical, oblate or prolate systems.

In our simulations, we found that these parameters do not vary significantly during the time evolution. In particular, the ratios $I_2/I_1$ and $I_3/I_1$ oscillate around the mean values $0.92$ and $0.83$, respectively.

A better indication on the triaxiality of a system is given by the {\it triaxiality parameter}, $T_{\rm tr}$, which is defined  as 
$T_{\rm tr} = (1-(I_2/I_1)^2)/(1-(I_3/I_1)^2)$.
According to \cite{Franx89}, $T_{\rm tr}=0$ corresponds to an oblate configuration, whereas $T_{\rm tr}=1$ is related to a prolate distribution. Values in the range $0<T_{\rm tr}<1$ characterize triaxial shaped systems.
In our numerical modelling, we found that the region within the inner 5 pc around the SMBH is clearly triaxial configuration, reaching the value $T_{\rm tr}\simeq 0.5$ after $\sim 0.3$ Gyr. 

On another side, the bottom panel of Fig. \ref{beta}, showing the mean line of sight velocity mapped within the inner $10$ pc, evidences that the GCs orbital evolution impinges a rotation along the 45 degree bisector of the x-y plane around the SMBH.

Such result is of particular interest in reference to the dynamics of stars around the SMBH hosted in the Andromeda galaxy nucleus, which has a mass of $M_{\rm SMBH} = (1.1-2.3)\times 10^8 \Ms$ \citep{Bender05} and a life-time $\gtrsim 100$ Myr.
Indeed, the Andomeda SMBH neighbourhood is characterised by the presence of a rotating, eccentric disk of stars  \citep{Lauer93,Tremaine95}, whose nature is still largely unclear.
A better understanding of the origin of this disky structures may help in shedding light on the SMBH growth history and process \citep{Hopkins10b}.

According to our present results, the infall of GCs seems to be inefficient in inducing the formation of such a configuration, at least when the central BH is very massive.

On the other hand, in our simulation all the GCs have initial distances to the galactic center larger than $\sim 10^2$ pc, and their disruption impinges only a mild rotation on the SMBH neighbourhood.

However, it is not clear whether a GC born close to the SMBH can give rise to a disk whose flatness survives for a time comparable to the estimeted age ($\sim 100$ Myr) of 
the Andromeda nuclear stellar disk.
In order to test such hypothesis, we made use of the data produced by \cite{ASCDS16}.

In particular, one of their sets of simulations was characterised by a galaxy model (and central SMBH) equal to the one presented in this work, and a GC moving at an initial distance of $50$ pc from the SMBH, i.e. smaller than those of the GCs in our numerical modelling, at varying GC initial eccentricity.

In the following, we refer to the simulation in which the GC moves on an orbit with $e_{\rm GC}\simeq 0.7$.

\begin{figure}
\centering
\includegraphics[width=9cm]{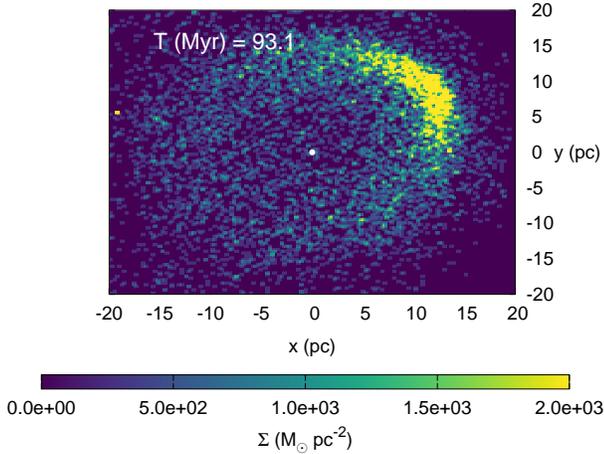}
\caption{Surface density map in the case of a single GC moving around the SMBH with initial distance $r_0=50$ pc and eccentricity $e\simeq 0.7$.}
\label{mapM31}
\end{figure}

Using these data, we show in Fig. \ref{mapM31} the surface density map of the cluster after $100$ Myr from the beginning of the simulation. It is quite evident the formation of a disky structure, with a projected radius extending up to $\sim 20$ pc, slightly larger than M31 disk, which extends up to $\sim 8$ pc.
Moreover, the bright pixel evident at $-20,5$ pc is due to a bunch of bounded stars, debris of the GC core. The total mass of this structure is $400\Ms$, and has an extension of $\sim 0.02$ pc.

This can be due to the GC model, which is based on a King density profile with $W_0=6$ and core radius $r_c=0.24$ pc. A more concentrated system could preserve a rounder shape on a longer time-scale, due to the deeper potential well.

These results suggest that the origin of disky structures in the immediate surroundings of a SMBH can be ascribed to the disruption of a relatively young GC born deep into the galactic nucleus.
It is worth noting that this scenario is complementary to another, suggested by some authors \citep{Tremaine95,hopkins10}, according to which the disk forms from a rotating gas cloud that undergoes subsequent star formation episodes.

\subsection{Tidal disruption events}
\label{TDE}

During the life of a galaxy evolution, some stars can move sufficiently close to the SMBH to be completely disrupted. During these tidal disruption events (TDEs, \cite{hills75}), part of the gas coming from the shattered star feeds the SMBH and can give rise to a detectable burst of X-rays. \cite{wang04} investigated the secular role of stellar dynamics around SMBHs, showing that the typical TDE rate for a heavy galactic nucleus is limited to a few $10^{-4}$ yr$^{-1}$ if the central SMBH exceeds $10^8\Ms$. In galaxies hosting SMBHs with masses $10^5-10^7\Ms$, recently \cite{stone17} pointed out that in dense, pre-existing, NSCs TDE rates can be in the range $10^{-5}-10^{-3}$ yr$^{-1}$, in dependence on the NSC properties. 

On the other hand, in galaxies characterised by a smooth, or cored, density profile hosting in their centers SMBH with mass above $10^8\Ms$ the TDE rate is generally limited in the range $10^{-6}-5\times 10^{-5}$ yr$^{-1}$  \citep{stone16}. Note that our galaxy model represents this kind of galaxies, as its density inner slope is small, $\gamma = 0.1$.

In this section, we investigate whether the GC debris, which accumulates around the SMBH, can enhance the TDE rate in a galaxy characterised by a cored density profile.

A star with radius $R_*$ and mass $M_*$ orbiting a SMBH undergoes a TDE if it approaches the SMBH closer than the so-called Roche radius

\begin{equation}
r_{\rm R} = \eta R_*\left(\frac{M_{\rm SMBH}}{M_*}\right)^{1/3},
\label{roche}
\end{equation}
with $\eta=0.8$ \citep{Merri13}.
As said above, these tidal disruption events are often followed by the emission of an X-rays flare with a time-scale of a few years. Nowadays, the detection of these strong signals represents a unique possibility to infer clues on the central SMBH mass and structure \citep{vinko15,kochanek16,yang16,metzger16}.

In order to express the Roche radius in terms of the SMBH and stellar properties, we recall here that $r_R$ is linked to the SMBH Schwarzschild's radius through the relation
 
\begin{equation}
\frac{r_{\rm R}}{r_{\rm S}} = 5.06\left(\frac{M_*}{\Ms}\right)^{-1/3}\left(\frac{M_{\rm SMBH}}{10^7 ~\Ms}\right)^{-2/3}\frac{R_*}{{\rm R}_\odot}.
\label{rRrS}
\end{equation}

Moreover, main sequence stars mass and radius are linked by a simple power-law 
\begin{equation}
\frac{R_*}{{\rm R}_\odot} = \alpha\left(\frac{M_*}{{\rm M}_\odot}\right)^\beta,
\end{equation}
with $\alpha$ and $\beta$ depending on the stellar mass, as shown in Table \ref{Radlaw} \citep{demircan91,gorda98}. Substituting into Eq. \ref{rRrS} we find
\begin{equation}
\frac{r_{\rm R}}{r_{\rm S}} = 5.06\left(\frac{M_{\rm SMBH}}{10^7 ~\Ms}\right)^{-2/3} \alpha\left(\frac{M_*}{\Ms}\right)^{-1/3+\beta}.
\label{rstar}
\end{equation}

Assuming $M_{\rm SMBH}=10^8$ M$_\odot$, Eq. \ref{rstar} implies that stars with mass smaller than $0.88$ M$_\odot$ undergo a direct plunge, and are wholly swallowed by the SMBH.
This clearly poses a limit to the number and type of TDEs and subsequent X-ray bursts.

\begin{table}
\caption{}
\centering{Parameters linking stellar radii and masses}
\begin{center}
\begin{tabular}{ccc}
\hline
                   & $\alpha$ & $\beta$ \\
\hline
 $M \leq 1.52 \Ms$  &   $1.09$   &  $0.969$  \\
 $M    > 1.52 \Ms$  &   $1.29$   &  $0.6035$ \\ 
\hline
\end{tabular}
\end{center}
\label{Radlaw}
\end{table}
 
For instance, under the assumption that the galaxy nucleus can be described by an isothermal sphere, for a SMBH with mass $\sim 10^8\Ms$ the expected rate of TDEs should be of the order of $10^{-4}\Ms$ yr$^{-1}$ \citep{Merri13}. Moreover, the TDE rate depends on the host galaxy density profile. Actually, while cored galaxies (inner slopes $\gamma<0.5$) have TDE rates in the range $10^{-6} - 5\times 10^{-5}\Ms$ yr$^{-1}$, steeper power-law galaxies are characterised by larger TDE rates.

So, the condition for a given star to lead to a TDE  is that $r_R/r_S \geq 1$ and the orbital pericentre, $r_p$, is $r_p \leq r_R$.

According to our model, the cumulative mass profile of the host galaxy is given by \citep{Deh93}
\begin{equation}
M(r) = M_g\left(\frac{r}{r+r_g}\right)^{3-\gamma}\simeq M_g\left(\frac{r}{r_g}\right)^{3-\gamma},
\label{M_r}
\end{equation}
the latter relation being valid for $r\ll r_g$, i.e. in the vicinity of the SMBH influence radius.
However, the GC infall changes the global density profile of the galaxy+GC system. As shown in Figure \ref{densglob}, the density profile gets steeper toward the center after $\gtrsim 200$ Myr, with a slope determined by the GCs orbital decay. 

In particular, the density distribution at $220$ Myr is sufficiently well fitted by a power-law

\begin{equation}
\rho(r) = \xi_0 r^{-\gamma_0},
\label{csi}
\end{equation}

with $\gamma_0=0.62\pm 0.06$ and $\xi_0 = 253 \pm 11 \Ms$ pc$^{-(3-\gamma_0)}$. Note that after the GC infall the value of $\gamma$ has increased from $0.1$ to $0.62$, a value which places our model slightly above that of the group of ``intermediate cusp'' galaxy (with $0.3<\gamma<0.5$, \cite{stone16}).

Integration gives the cumulative, inner, mass distribution

\begin{equation}
M_f(r) = 4\pi\int_0^r  r^2 \rho(r)dr = \frac{4\pi \xi_0}{3-\gamma_0}r^{3-\gamma_0}.
\label{M_0}
\end{equation}

leading to a ratio between the final and initial mass radial profiles which is
\begin{equation}
\frac{M_f(r)}{M(r)} = \frac{4\pi \xi_0 }{(3-\gamma_0)M_g}\frac{r^{\gamma-\gamma_0}}{r_g^{\gamma -3}}.
\label{mur}
\end{equation}

Substituting in Eq. \ref{mur} the relevant quantities discussed above we find that the above mass ratio attains a value of $\sim 2\times 10^3$ at $r=10^{-3}$ pc.
Due to that the TDE rate is proportional to the galaxy density \citep{wang04}, we expect that $\dot{N}_{\rm TDE}$ after the GC infall would thus increase by a factor $\sim 10^3$.
As said above, a correct analysis of possible TDEs must account only for stars moving on orbits whose pericentres fall below $r_{\rm R}$ and mass greater than $0.88$ M$_\odot$.

As discussed in Section \ref{dftd}, in our model the SMBH influence radius is $\gtrsim 10$ pc, significantly larger than $r_{\rm R}$. 

I we consider stars formerly belonging to GCs and which are now buzzing around the SMBH,  we can evaluate their orbital pericentres, $r_p$, in the two-body approximation

\begin{equation}
r_p = (1-e_*) \frac{1}{2/r-v_*^2/(GM_{\rm SMBH})};
\label{peric}
\end{equation}

This approximation is valid because these stars are confined in the SMBH influence sphere whose radius is $\sim 13$. For this reason
we can substitute $v_*^2/(GM_{\rm SMBH})$ with $r_{\rm inf}$ in the equation above to obtain

\begin{equation}
r_p = \frac{1-e_*}{2-r/r_{\rm inf}}r.
\label{rperi}
\end{equation}
 
all the stars that will likely undergo a TDE are those moving in the inner region around the SMBH, at distances below $10^{-3}$ pc. These stars are debris of the dissolved population of GCs, and their orbital apocentre hardly can exceed the SMBH influence radius, which is $\sim 13$ pc by the end of our simulation. Hence, we can substitute $\sigma_g^2/(GM_{\rm SMBH}$ with $r_{\rm inf}$ in the equation above to obtain

\begin{equation}
r_p = \frac{1-e_*}{2-r/r_{\rm inf}}r.
\label{rperi}
\end{equation}

To have a TDE, the condition $r_p\leq r_{\rm R}$ leads to

\begin{equation}
r_p \leq \frac{2r_{\rm R}}{1-e_*+r_{\rm R}/r_{\rm inf}}\simeq \frac{r_{\rm R}}{1-e_*}.
\label{tde}
\end{equation}

Hence, even for very eccentric orbit, the star distance to the SMBH must be small (less than $10r_{\rm R}$ for $e_*=0.9$), too small to be resolved with our simulation, despite its high level of detail.

\begin{figure}
\centering
\includegraphics[width=8cm]{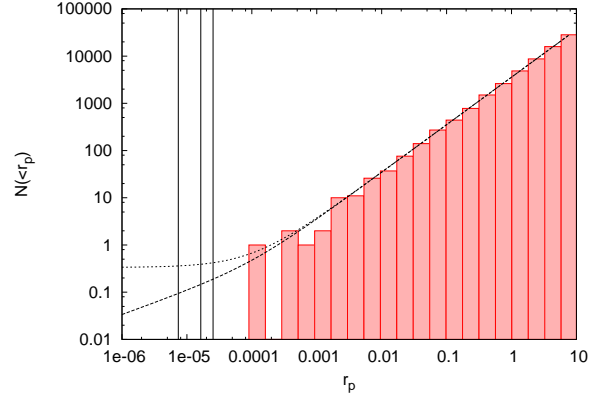}
\caption{Cumulative distribution of the stars pericentre. From left to right, the black vertical lines represent the Roche radius for a $0.5$, $2$ and $10 \Ms$ star. the two black curves represent the fitting functions described in Eqs. \ref{f1} and \ref{f2}.
}
\label{TDEdist}
\end{figure}

For a Sun-like star, the pericentre threshold to have a TDE corresponds, roughly, to $10^{-5}$ pc, increasing up to a few $10^{-4}$ in the case of a $100 \Ms$ star,  at least one order of magnitude smaller than our simulation resolution.
Nevertheless, we can use our results to extrapolate the number of stars expected to undergo a TDE in a real galaxy. 

In Fig. \ref{TDEdist} the cumulative distribution $N(<r_p)$ of the stars pericentre is shown. In order to give an estimate of the total number of stars that can be disrupted by the SMBH tidal force, we should extrapolate this distribution toward small values of $r_p$. 

Due to the limited resolution of our $N$-body simulation, below $10^{-3}$ pc, we cannot state clearly whether the low-end tail of $N_{r_p}$ tends to a constant value or rapidly drops to zero, giving a huge uncertainty in the extrapolation procedure.
Due to this, we decided to search for two different fitting functions, able to reproduce the two extreme cases of (1) an $N_{r_p}$ which drops rapidly to 0 at decreasing pericentre, on one side, and that of (2) an $N_{r_p}$  which tends to a constant at small values of $r_p$.

The rapidly decreasing function (1), named $f_1$, is defined as
\begin{equation}
f_1(r_p) = kc(ar_p+1)^{b}\sqrt{r_p},
\label{f1}
\end{equation}

 while the other function, $f_2$, is given by
acknowledge
\begin{equation}
f_2(r_p) = kc(ar_p+1)^{b},
\label{f2}
\end{equation}
In both equations, shown in Figure \ref{remndistri}, $k=1/N_{\rm GCS}$ represents the inverse of the number of particles used to represent all the GCs. Moreover, we set $1/a=10^{-4}$ pc, which is the length scale below which our resolution in $N_{r_p}$ loses quality.

Once the two functions have been selected, we used the non-linear least-squares (NLLS) Marquardt-Levenberg algorithm implemented in the analysis tool \texttt{GNUPLOT}, to provide the set of parameters that describe at best the $N_{r_p}$.

The value of the fitting parameters are resumed in Table \ref{fit}.

\begin{table}
\caption{Parameters of the $N_{r_p}$ fitting functions}
\begin{center}
\begin{tabular}{cccc}
\hline
	& $a$ & $b$ & $c$\\
\hline
$f_1(r_p)$ & $10^4$ & $0.507 \pm 0.003$& $34 \pm 1$\\
$f_2(r_p)$ & $10^4$ & $1.007 \pm 0.003$& $0.33 \pm 0.01$ \\
\hline     
\end{tabular}
\end{center}
\label{fit}
\end{table}
 
Among all the stars, those having an angular momentum smaller than a limiting value, called loss-cone angular momentum $L_{\rm LC} = \sqrt{GM_{\rm SMBH}r_R}$, will undergo a TDE over a relaxation time \citep{rees88,Merri13}.
Following \cite{wang04}, only stars having pericentres smaller than a critical radius, $r_{\rm crit}$, have orbital properties that can cause the deflection of stars into the loss-cone regime. According to their definition, the TDE rate is given by the ratio between the number of stars having a pericentre smaller than $r_{\rm crit}$ and the local relaxation time calculated at this radius
\begin{equation}
\dot{N}_{\rm TDE}(<r_{\rm crit}) = \frac{N_{\rm TDE}(<r_{\rm crit})}{T_r(r<r_{\rm crit})},
\label{TDErate}
\end{equation}
where $N_{\rm TDE}(<r_{\rm crit})$ represent the number of stars having pericentre smaller than $r_{\rm crit}$
, and where we used the 2-body relaxation time-scale as defined by \cite{spitzer58}
\begin{equation}
T_r(r) = \frac{\sqrt{2}\sigma_g(r)^3}{\pi G^2 m_* \rho(r) \ln\Lambda}.
\end{equation}

In order to calculate $r_{\rm crit}$, we must set the so-called ``{\it loss-cone angle}'', $\theta_{\rm LC}$
\begin{equation}
\theta_{\rm LC}^2=r_R/r_{\rm crit},
\end{equation}
to be equal to the angle, $\theta_d$, by which a star orbit is deflected into the loss-cone, which is the ratio between the star orbital period and the local relaxation time-scale
\begin{equation}
\theta_d^2 = \frac{\sqrt{r_{\rm crit}^3/GM_{\rm SMBH}}}{T_r(r_{\rm crit})}.
\end{equation}

The condition $\theta_d = \theta_{\rm LC}$ implies
\begin{equation}
r_{\rm crit}^{5/2-\gamma_0} =  
\frac{
\sqrt{2} \eta \alpha M_{\rm SMBH}^{5/6}m_*^{-4/3+\beta} \sigma_g^3
}
{
\pi \ln\Lambda \xi_0
}.
\end{equation}

Given the dependence of the Roche radius on the star mass (see Eq. \ref{rRrS}), to get the fraction of stars which likely undergo to TDE
we need to know the GC mass function.

Assuming a Salpeter initial mass function \citep{Salpeter55},
the fraction of stars having a mass greater than the limiting value above which the pericentre distance to the SMBH is smaller than its Roche radius is given by

\begin{equation}
\nu(r_p<r_R,\tau=0) = \frac{m_M^{1-s}-m_p^{1-s}}{m_M^{1-s}-m_m^{1-s}},
\label{nu}
\end{equation}

where $m_p$ is the star mass which gives $r_R=r_p$, $s = 2.35$ and $m_{m} = 0.1\Ms$ is the minimum mass and $m_{M} = 100\Ms$ the  maximum mass in the IMF (age zero, i.e. $\tau=0$). 

To account for the time evolution of the MF due to star mass loss, to evaluate the proper value of $\nu$, we followed the procedure described in \cite{AS16}, which makes use of the stellar evolution code SSE \citep{hurley2000}.
Following the time evolution of the population of stars with masses in the range $m_{m}-m_{M}$ up to $1$ Gyr we found that the evolved MF shows at any time a steep decline at masses above $m_e$ defined as the mass of stars ending their H burning phase at that time (it represents the minimum mass of unevolved stars at a given age).

We report in Table \ref{me} some values of $m_e$ at different times, highlighting the fraction of stars heavier than $1\Ms$. We gave estimate assuming for the GCs either a solar metallicity, $Z_\odot =0.02$, or a low value, $Z_\odot =0.0004$, typical of old GCs in the MW.

Note that, in the range of ages of Table \ref{me}, $m_e$ is such to give a Roche radius greater than the SMBH $r_S$.

\begin{figure}
\includegraphics[width=8cm]{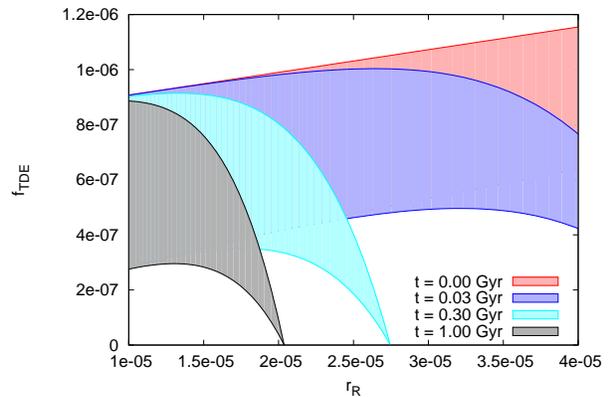}\\
\caption{Total fraction of stars with Roche radius larger than their pericentre, $r_p<r_R$. The lower limit of the shaded region is obtained assuming in our calculation $f_1(r_p)$, while the upper limit is obtained assuming $f_2(r_p)$.}
\label{TDEfrac}
\end{figure}

Finally, in our simulation the fraction of stars which may give origin to TDEs is given by

\begin{equation}
f_{\rm TDE}(< r_p) = f_i(r_p)\nu(r_R>r_p,\tau),
\label{TDEnumber}
\end{equation}

where $f_i, i=1,2$ are the functions in Eqs. \ref{f1} and \ref{f2}.
The choice $f_1(r_R)$ minimizes the fraction of TDEs, while $f_2(r_R)$ maximizes it. In Fig. \ref{TDEfrac} we show how $f_{\rm TDE}$ varies as a function of the pericentre distance 

to the galactic center assuming different values of the GC age.

\begin{table}
\caption{}
\begin{center}
\begin{tabular}{cccc}
\hline
$T$ & $m_e$($Z_\odot = 0.02$) & $m_e$($Z_\odot = 0.0004$) & $n(m>1\Ms)$\\
Gyr & $\Ms$& $\Ms$& $(\%)$\\
\hline
0   & 100 & 100 & 4.5\\
0.01& 17  & 19  & 4.4\\
0.03& 9.2 & 9.3 & 4.3\\
0.3 & 4   & 4.5 & 3.8\\
1   & 2.5 & 2.0 & 3.2\\
\hline
\end{tabular}
\end{center}
\begin{tablenotes}
\item Col. 1: time. Col. 2: minimum unevolved mass at solar metallicity. Col 3.:  minimum unevolved mass at metallicity $Z=0.0004$. Col. 4: fraction of stars with mass larger than $1\Ms$.

\end{tablenotes}
\label{me}
\end{table}

Therefore, the TDE rate can be evaluated substituting $r_{\rm crit}$ in Equation \ref{TDErate} and using Equation \ref{TDEnumber}
\begin{equation}
\dot{N}_{\rm TDE}(<r_{\rm crit}) = 
\frac{f_{\rm TDE}(<r_{\rm crit})N_*} {T_r(r_{\rm crit})} .
\end{equation}

In our calculations we considered a total number of stars $N_* \simeq 2\times 10^{11}$, a velocity dispersion $\sigma_g(r)\sim 100$ km s$^{-1}$, as evaluated in our simulation at distances below $1$ pc from the SMBH, and a density at $r\sim 10^{-5}$ pc given by $\rho(r) \simeq 3\times 10^5 \Ms$ pc$^{-3}$, as evaluated by Eq. \ref{csi}. Note that $r_p>10^{-5}$ pc is the minimum distance above which stars pericentre is larger than the SMBH Schwarzschild radius, thus representing the spatial region where the probability to have a TDE is maximized.

By substituting the relevant quantities, we found a TDE rate $\dot{N}_{\rm TDE} = 1.9\times 10^{-4}$ yr$^{-1}$.
We note here that galaxies with $\gamma$ in the range $0.3-0.8$ and SMBH with masses $\sim 10^8\Ms$ are characterised by a TDE rate $\simeq (1.9-4.5)\times 10^{-5}$ yr$^{-1}$ (cfr. Table 1 in \cite{stone16}), which is an order of magnitude lower than what we found in our simulations, which is characterised by $\gamma\sim 0.6$. 
Additionally, it seems that the mass accumulated around the SMBH in form of stellar debris can actually determine an enhancement in the TDE rate. 
Anyway, the above result depends of various assumptions, as discussed in the following subsection.
This is particularly interesting in the context of the recent discovery of a large TDE rate in the E+A galaxy NGC3156. E+A galaxies are elliptical galaxies which underwent a starburst formation episode $0.1-1$ Gyr ago and are often populated by a population of young massive clusters \citep{yang04}.
NGC3516 is characterised by an SMBH significantly lighter than that of our model, with a mass $0.9-2.7 \times 10^6\Ms$ \citep{stone16b}.
Assuming that the TDE rate is linked to the SMBH mass through a power-law \citep{stone16}, we can rescale our results to the SMBH mass of NGC3516 getting a TDE rate $\dot{N}_{\rm TDE} = (0.6-1.1)\times 10^{-3}$ yr$^{-1}$, compatible with the observed estimated value, thus suggesting that the infall and merger of a population of young star clusters can enhance significantly the TDE rate. 

\subsection*{Caveats of TDEs calculation}
The results obtained for TDE rates depend strongly on the galaxy model and the GCs masses and orbital properties.
In our calculations, the main parameters that affect the TDE rate are: i) the GC position and orbital velocity distributions and ii) the GC mass function and total number.

Regarding the first point, we assumed that the GC and host galaxy distributions does not differ significantly. 
Such simple assumption allows to explain several properties of galactic centres, for instance the masses and sizes of NCs  \citep{antonini13,ASCD14b,gnedin14}, the $\gamma$-ray flux observed in the Milky Way central regions \citep{brandt15} and the presence of an old population of red giant stars in the Milky Way NC \citep{min16}. 
A GCs density profile steeper than the galaxy would lead to too large NCs, due to the high efficiency of df in the galaxy centre vicinity, while shallower density profiles would lead to too small NCs, as in this case most of the GCs can be disrupted by the galactic tidal forces.

Regarding the second point, the total mass in massive ($M>10^5\Ms$) star clusters in a heavy galaxy is roughly $0.1-1\%$ of the galaxy mass \citep{harris10}. 
Assuming that GC masses range in between $(0.3-2)\times 10^6\Ms$ and that they are distributed according to the background galaxy distribution, we expect, for a $10^{11}\Ms$ galaxy, between 8 and 80 GCs within $100$ pc from the SMBH. Therefore, it seems that $\gtrsim 40$ GCs is a reasonable choice, given our current knowledge of galaxy formation and evolution. 
Our choice of the GC minimum and maximum mass allowed is dictated by the computational need of having a sufficient number of particles to represent the smallest clusters. However, such choice is well supported by previous theoretical works which tackled similar problems (see for instance \cite{antonini13}, \cite{ASCD14b} and \cite{gnedin14}).
Moreover, our simulation shows that only 16 GCs, out of the 42, reach the inner 10 pc (see Figure \ref{GCdist}) and undergo a strong encounter with the SMBH. The remaining clusters are subjected to the strong action of the galactic tidal field, and never approach close to the galactic centre, thus they do not contribute to the density enhancement that leads to the increase in the TDE rate.

\section{Conclusions}
\label{conclusions}

In this paper we modelled the evolution of 42 GCs moving in the nucleus of a massive galaxy hosting a $10^8$ M$_\odot$ SMBH at its centre. 
This was done self-consistently, by mean of a numerical representation at an unprecedented level of detail in this framework.
The simulation outcomes show that tidal torques due to the combined effect of the background galaxy and SMBH shape the properties of the SMBH surroundings, eroding the infalling GCs and quenching the formation of a dense nucleus.

Indeed, although the GCs orbital decay leads to an increase in the galactic spatial density, it is not sufficiently efficient to produce a clear enhancement of the central surface density profile, which is widely used to infer the presence of an NSC. 

Using the output of our simulations, we also investigated the role of SMBH-GCs interactions in determining stellar TDEs rate.

The main outcomes of this paper can be summarized as follows:

\begin{itemize}

\item[i)] we found that tidal forces and dynamical friction acts rapidly in determining the GC evolution, suggesting that GC-SMBH interactions can shape a galactic nucleus on time-scales smaller than $1$ Gyr;

\item[ii)] the tidal torques induced by the central SMBH on its surroundings are such to shatter most of the GCs that approach the galactic central region, causing an inefficient deposit of mass, in form of GC debris, around the SMBH. This provide a reliable explanation for the absence of NSC in galaxies hosting SMBHs heavier than $10^8\Ms$, as outlined in other recent papers \citep{ASCD14b,antonini13,ASCDS16};

\item[iii)] the GCs debris accumulated around the SMBH impinge a clear kinematic fingerprint on the galactic nucleus. In  particular, our results show that the innermost region around the SMBH is characterised by a flattened configuration, strongly triaxial and weakly rotating;

\item[iv)] the interaction among the SMBH, the stellar field and the infalling 42 GCs shapes significantly the GCs mass distribution. In particular, GC mass loss induced by tidal forces leads to a sub-population of GCs with masses below $3\times 10^4\Ms$, moving at $\sim 50-100$ pc from the galactic centre. Above such limiting value, their mass distribution is well-described by a power-law, characterised by a slope $\simeq -0.8$;

\item[v)] comparing our results with \cite{ASCDS16}, we note here that if a massive cluster forms in the SMBH vicinity, its disruption can lead to the formation of a disky structure with a life-time $\simeq 100$ Myr. This provide a further explanation for the origin of dense stellar disks around SMBHs with masses around $10^8\Ms$, as observed, for instance, in the M31 galaxy \citep{Lauer93,Tremaine95};

\item[vi)] using the huge amount of data produced by our simulation, we investigated whether GCs-SMBH interactions can enhance the probability of flares from tidally destroyed stars belonging to GCs passing by the SMBH. We found a TDE rate of $\sim 1.9\times 10^{-4}$ yr$^{-1}$, a value significantly larger than what expected for galaxies characterised by a similarly steep density profile \citep{stone16}, which finds an interesting agreement in recent observations of several E+A galaxies \citep{stone16b}.

\end{itemize}

\section{Acknowledgements}
The authors acknowledge the referee, whose helpful comments and suggestions allowed to improve an earlier version of the manuscript.
MAS acknowledges Sapienza, University of Rome, which funded the research program ``MEGaN: modelling the evolution of galactic nuclei'' via the grant 52/2015, and the Sonderforschungsbereich SFB 881 "The Milky Way System" of the German Research Foundation (DFG) for the financial support provided. MAS also thanks the Zentrum fur Astronomie - Astronomisches Rechen-Institut of Heidelberg for the hospitality during the development of part of this research.  

\clearpage

\footnotesize{
\bibliographystyle{mn2e}
\bibliography{ASetal2015}

\newcommand{\noop}[1]{}
\begin{thebibliography}{}

\bibitem[\protect\citeauthoryear{{Aharon} \& {Perets}}{{Aharon} \&
  {Perets}}{2015}]{aharon15}
{Aharon} D.,  {Perets} H.~B.,  2015, \apj, 799, 185

\bibitem[\protect\citeauthoryear{{Antonini}}{{Antonini}}{2013}]{antonini13}
{Antonini} F.,  2013, \apj, 763, 62

\bibitem[\protect\citeauthoryear{{Antonini}, {Barausse} \& {Silk}}{{Antonini}
  et~al.}{2015}]{Antonini15}
{Antonini} F.,  {Barausse} E.,    {Silk} J.,  2015, \apj, 812, 72

\bibitem[\protect\citeauthoryear{{Antonini}, {Capuzzo-Dolcetta},
  {Mastrobuono-Battisti} \& {Merritt}}{{Antonini} et~al.}{2012}]{AMB}
{Antonini} F.,  {Capuzzo-Dolcetta} R.,  {Mastrobuono-Battisti} A.,    {Merritt}
  D.,  2012, \apj, 750, 111

\bibitem[\protect\citeauthoryear{{Antonini} \& {Merritt}}{{Antonini} \&
  {Merritt}}{2012}]{AntMer12}
{Antonini} F.,  {Merritt} D.,  2012, \apj, 745, 83

\bibitem[\protect\citeauthoryear{{Arca-Sedda}}{{Arca-Sedda}}{2016}]{AS16}
{Arca-Sedda} M.,  2016, \mnras, 455, 35

\bibitem[\protect\citeauthoryear{{Arca-Sedda} \&
  {Capuzzo-Dolcetta}}{{Arca-Sedda} \& {Capuzzo-Dolcetta}}{2014a}]{ASCD14a}
{Arca-Sedda} M.,  {Capuzzo-Dolcetta} R.,  2014a, \apj, 785, 51

\bibitem[\protect\citeauthoryear{{Arca-Sedda} \&
  {Capuzzo-Dolcetta}}{{Arca-Sedda} \& {Capuzzo-Dolcetta}}{2014b}]{ASCD14b}
{Arca-Sedda} M.,  {Capuzzo-Dolcetta} R.,  2014b, \mnras, 444, 3738

\bibitem[\protect\citeauthoryear{{Arca-Sedda} \&
  {Capuzzo-Dolcetta}}{{Arca-Sedda} \& {Capuzzo-Dolcetta}}{2016}]{ASCD16a}
{Arca-Sedda} M.,  {Capuzzo-Dolcetta} R.,  2016, \mnras, 461, 4335

\bibitem[\protect\citeauthoryear{{Arca-Sedda} \&
  {Capuzzo-Dolcetta}}{{Arca-Sedda} \& {Capuzzo-Dolcetta}}{2017}]{ASCD16b}
{Arca-Sedda} M.,  {Capuzzo-Dolcetta} R.,  2017, \mnras, 464

\bibitem[\protect\citeauthoryear{{Arca-Sedda}, {Capuzzo-Dolcetta}, {Antonini}
  \& {Seth}}{{Arca-Sedda} et~al.}{2015}]{ASCD15He}
{Arca-Sedda} M.,  {Capuzzo-Dolcetta} R.,  {Antonini} F.,    {Seth} A.,  2015,
  \apj, 806, 220

\bibitem[\protect\citeauthoryear{{Arca-Sedda}, {Capuzzo-Dolcetta} \&
  {Spera}}{{Arca-Sedda} et~al.}{2016}]{ASCDS16}
{Arca-Sedda} M.,  {Capuzzo-Dolcetta} R.,    {Spera} M.,  2016, \mnras, 456,
  2457

\bibitem[\protect\citeauthoryear{{Bekki}}{{Bekki}}{2007}]{bekki07}
{Bekki} K.,  2007, PASA, 24, 77

\bibitem[\protect\citeauthoryear{{Bekki} \& {Graham}}{{Bekki} \&
  {Graham}}{2010}]{bekki10}
{Bekki} K.,  {Graham} A.~W.,  2010, \apjl, 714, L313

\bibitem[\protect\citeauthoryear{{Bender}, {Kormendy}, {Bower}, {Green},
  {Thomas}, {Danks}, {Gull}, {Hutchings}, {Joseph}, {Kaiser}, {Lauer},
  {Nelson}, {Richstone}, {Weistrop} \& {Woodgate}}{{Bender}
  et~al.}{2005}]{Bender05}
{Bender} R.,  {Kormendy} J.,  {Bower} G.,  {Green} R.,  {Thomas} J.,  {Danks}
  A.~C.,  {Gull} T.,  {Hutchings} J.~B.,  {Joseph} C.~L.,  {Kaiser} M.~E.,
  {Lauer} T.~R.,  {Nelson} C.~H.,  {Richstone} D.,  {Weistrop} D.,
  {Woodgate} B.,  2005, \apj, 631, 280

\bibitem[\protect\citeauthoryear{{B{\"o}ker}, {Laine}, {van der Marel},
  {Sarzi}, {Rix}, {Ho} \& {Shields}}{{B{\"o}ker} et~al.}{2002}]{boker02}
{B{\"o}ker} T.,  {Laine} S.,  {van der Marel} R.~P.,  {Sarzi} M.,  {Rix} H.-W.,
   {Ho} L.~C.,    {Shields} J.~C.,  2002, \aj, 123, 1389

\bibitem[\protect\citeauthoryear{{B{\"o}ker}, {Sarzi}, {McLaughlin}, {van der
  Marel}, {Rix}, {Ho} \& {Shields}}{{B{\"o}ker} et~al.}{2004}]{boker04}
{B{\"o}ker} T.,  {Sarzi} M.,  {McLaughlin} D.~E.,  {van der Marel} R.~P.,
  {Rix} H.-W.,  {Ho} L.~C.,    {Shields} J.~C.,  2004, \aj, 127, 105

\bibitem[\protect\citeauthoryear{{Bonfini} \& {Graham}}{{Bonfini} \&
  {Graham}}{2016}]{bonfini16}
{Bonfini} P.,  {Graham} A.~W.,  2016, \apj, 829, 81

\bibitem[\protect\citeauthoryear{{Brandt} \& {Kocsis}}{{Brandt} \&
  {Kocsis}}{2015}]{brandt15}
{Brandt} T.~D.,  {Kocsis} B.,  2015, \apj, 812, 15

\bibitem[\protect\citeauthoryear{{Capuzzo-Dolcetta}}{{Capuzzo-Dolcetta}}{1993}%
]{Dolc93}
{Capuzzo-Dolcetta} R.,  1993, \apj, 415, 616

\bibitem[\protect\citeauthoryear{{Capuzzo-Dolcetta} \&
  {Fragione}}{{Capuzzo-Dolcetta} \& {Fragione}}{2015}]{fragione15}
{Capuzzo-Dolcetta} R.,  {Fragione} G.,  2015, \mnras, 454, 2677

\bibitem[\protect\citeauthoryear{{Capuzzo-Dolcetta} \&
  {Miocchi}}{{Capuzzo-Dolcetta} \& {Miocchi}}{2008}]{DoMioA}
{Capuzzo-Dolcetta} R.,  {Miocchi} P.,  2008, \mnras, 388, L69

\bibitem[\protect\citeauthoryear{{Capuzzo-Dolcetta}, {Spera} \&
  {Punzo}}{{Capuzzo-Dolcetta} et~al.}{2013}]{Spera}
{Capuzzo-Dolcetta} R.,  {Spera} M.,    {Punzo} D.,  2013, Journal of
  Computational Physics, 236, 580

\bibitem[\protect\citeauthoryear{{Chandar}, {Leitherer}, {Tremonti} \&
  {Calzetti}}{{Chandar} et~al.}{2003}]{chandar03}
{Chandar} R.,  {Leitherer} C.,  {Tremonti} C.,    {Calzetti} D.,  2003, \apj,
  586, 939

\bibitem[\protect\citeauthoryear{{Chandrasekhar}}{{Chandrasekhar}}{1943}]{Cha4%
3I}
{Chandrasekhar} S.,  1943, \apj, 97, 255

\bibitem[\protect\citeauthoryear{{C{\^o}t{\'e}}, {Piatek}, {Ferrarese},
  {Jord{\'a}n}, {Merritt}, {Peng}, {Ha{\c s}egan}, {Blakeslee}, {Mei}, {West},
  {Milosavljevi{\'c}} \& {Tonry}}{{C{\^o}t{\'e}} et~al.}{2006}]{cote06}
{C{\^o}t{\'e}} P.,  {Piatek} S.,  {Ferrarese} L.,  {Jord{\'a}n} A.,  {Merritt}
  D.,  {Peng} E.~W.,  {Ha{\c s}egan} M.,  {Blakeslee} J.~P.,  {Mei} S.,  {West}
  M.~J.,  {Milosavljevi{\'c}} M.,    {Tonry} J.~L.,  2006, ApJS, 165, 57

\bibitem[\protect\citeauthoryear{{Davies}, {Miller} \& {Bellovary}}{{Davies}
  et~al.}{2011}]{davies11}
{Davies} M.~B.,  {Miller} M.~C.,    {Bellovary} J.~M.,  2011, \apjl, 740, L42

\bibitem[\protect\citeauthoryear{{Dehnen}}{{Dehnen}}{1993}]{Deh93}
{Dehnen} W.,  1993, \mnras, 265, 250

\bibitem[\protect\citeauthoryear{{Demircan} \& {Kahraman}}{{Demircan} \&
  {Kahraman}}{1991}]{demircan91}
{Demircan} O.,  {Kahraman} G.,  1991, \apss, 181, 313

\bibitem[\protect\citeauthoryear{{den Brok}, {Peletier}, {Seth}, {Balcells},
  {Dominguez}, {Graham}, {Carter}, {Erwin}, {Ferguson} \& {et al .}}{{den Brok}
  et~al.}{2014}]{denB}
{den Brok} M.,  {Peletier} R.~F.,  {Seth} A.,  {Balcells} M.,  {Dominguez} L.,
  {Graham} A.~W.,  {Carter} D.,  {Erwin} P.,  {Ferguson} H.~C.,    {et al .}
  2014, \mnras, 445, 2385

\bibitem[\protect\citeauthoryear{Donnari, Arca-Sedda \& Graham}{Donnari
  et~al.}{2017}]{donnari17}
Donnari M.,  Arca-Sedda M.,    Graham A.~W.,  2017, in prep.

\bibitem[\protect\citeauthoryear{{Dosopoulou} \& {Antonini}}{{Dosopoulou} \&
  {Antonini}}{2016}]{dosopoulou17}
{Dosopoulou} F.,  {Antonini} F.,  2016, ArXiv e-prints

\bibitem[\protect\citeauthoryear{{Dressler} \& {Gunn}}{{Dressler} \&
  {Gunn}}{1983}]{dressler83}
{Dressler} A.,  {Gunn} J.~E.,  1983, \apj, 270, 7

\bibitem[\protect\citeauthoryear{{Emsellem}}{{Emsellem}}{2013}]{Ems13}
{Emsellem} E.,  2013, \mnras, 433, 1862

\bibitem[\protect\citeauthoryear{{Erwin} \& {Gadotti}}{{Erwin} \&
  {Gadotti}}{2012}]{ERWGD}
{Erwin} P.,  {Gadotti} D.~A.,  2012, Advances in Astronomy, 2012

\bibitem[\protect\citeauthoryear{{Ferrarese}, {C{\^o}t{\'e}}, {Dalla
  Bont{\`a}}, {Peng}, {Merritt}, {Jord{\'a}n}, {Blakeslee}, {Ha{\c s}egan},
  {Mei}, {Piatek}, {Tonry} \& {West}}{{Ferrarese} et~al.}{2006}]{frrs}
{Ferrarese} L.,  {C{\^o}t{\'e}} P.,  {Dalla Bont{\`a}} E.,  {Peng} E.~W.,
  {Merritt} D.,  {Jord{\'a}n} A.,  {Blakeslee} J.~P.,  {Ha{\c s}egan} M.,
  {Mei} S.,  {Piatek} S.,  {Tonry} J.~L.,    {West} M.~J.,  2006, \apjl, 644,
  L21

\bibitem[\protect\citeauthoryear{{Fragione}, {Capuzzo-Dolcetta} \&
  {Kroupa}}{{Fragione} et~al.}{2017}]{fragione17}
{Fragione} G.,  {Capuzzo-Dolcetta} R.,    {Kroupa} P.,  2017, \mnras, 467, 451

\bibitem[\protect\citeauthoryear{{Franx}, {Illingworth} \& {Heckman}}{{Franx}
  et~al.}{1989}]{Franx89}
{Franx} M.,  {Illingworth} G.,    {Heckman} T.,  1989, \aj, 98, 538

\bibitem[\protect\citeauthoryear{{Georgiev} \& {B{\"o}ker}}{{Georgiev} \&
  {B{\"o}ker}}{2014}]{georgiev14}
{Georgiev} I.~Y.,  {B{\"o}ker} T.,  2014, \mnras, 441, 3570

\bibitem[\protect\citeauthoryear{{Georgiev}, {B{\"o}ker}, {Leigh},
  {L{\"u}tzgendorf} \& {Neumayer}}{{Georgiev} et~al.}{2016}]{georgiev16}
{Georgiev} I.~Y.,  {B{\"o}ker} T.,  {Leigh} N.,  {L{\"u}tzgendorf} N.,
  {Neumayer} N.,  2016, \mnras, 457, 2122

\bibitem[\protect\citeauthoryear{{Gnedin}, {Ostriker} \& {Tremaine}}{{Gnedin}
  et~al.}{2014}]{gnedin14}
{Gnedin} O.~Y.,  {Ostriker} J.~P.,    {Tremaine} S.,  2014, \apj, 785, 71

\bibitem[\protect\citeauthoryear{{Gorda} \& {Svechnikov}}{{Gorda} \&
  {Svechnikov}}{1998}]{gorda98}
{Gorda} S.~Y.,  {Svechnikov} M.~A.,  1998, Astronomy Reports, 42, 793

\bibitem[\protect\citeauthoryear{{Graham}}{{Graham}}{2012}]{graham12}
{Graham} A.~W.,  2012, \mnras, 422, 1586

\bibitem[\protect\citeauthoryear{{Graham}, {Onken}, {Athanassoula} \&
  {Combes}}{{Graham} et~al.}{2011}]{Graham11}
{Graham} A.~W.,  {Onken} C.~A.,  {Athanassoula} E.,    {Combes} F.,  2011,
  \mnras, 412, 2211

\bibitem[\protect\citeauthoryear{{Graham} \& {Spitler}}{{Graham} \&
  {Spitler}}{2009}]{graham09}
{Graham} A.~W.,  {Spitler} L.~R.,  2009, \mnras, 397, 2148

\bibitem[\protect\citeauthoryear{{Harris}}{{Harris}}{2010}]{harris10}
{Harris} W.~E.,  2010, Philosophical Transactions of the Royal Society of
  London Series A, 368, 889

\bibitem[\protect\citeauthoryear{{Hills}}{{Hills}}{1975}]{hills75}
{Hills} J.~G.,  1975, \nat, 254, 295

\bibitem[\protect\citeauthoryear{{Hopkins} \& {Quataert}}{{Hopkins} \&
  {Quataert}}{2010a}]{hopkins10}
{Hopkins} P.~F.,  {Quataert} E.,  2010a, \mnras, 407, 1529

\bibitem[\protect\citeauthoryear{{Hopkins} \& {Quataert}}{{Hopkins} \&
  {Quataert}}{2010b}]{Hopkins10b}
{Hopkins} P.~F.,  {Quataert} E.,  2010b, \mnras, 405, L41

\bibitem[\protect\citeauthoryear{{Hurley}, {Pols} \& {Tout}}{{Hurley}
  et~al.}{2000}]{hurley2000}
{Hurley} J.~R.,  {Pols} O.~R.,    {Tout} C.~A.,  2000, \mnras, 315, 543

\bibitem[\protect\citeauthoryear{{King}}{{King}}{2003}]{King03}
{King} A.,  2003, \apjl, 596, L27

\bibitem[\protect\citeauthoryear{{King}}{{King}}{2005}]{King05}
{King} A.,  2005, \apjl, 635, L121

\bibitem[\protect\citeauthoryear{{King}}{{King}}{1962}]{king62}
{King} I.,  1962, \aj, 67, 471

\bibitem[\protect\citeauthoryear{{King}}{{King}}{1966}]{King}
{King} I.~R.,  1966, \aj, 71, 64

\bibitem[\protect\citeauthoryear{{Kochanek}}{{Kochanek}}{2016}]{kochanek16}
{Kochanek} C.~S.,  2016, \mnras, 461, 371

\bibitem[\protect\citeauthoryear{{Kormendy} \& {Ho}}{{Kormendy} \&
  {Ho}}{2013}]{Kormendy13}
{Kormendy} J.,  {Ho} L.~C.,  2013, \araa, 51, 511

\bibitem[\protect\citeauthoryear{{Lauer}, {Faber}, {Groth}, {Shaya},
  {Campbell}, {Code}, {Currie}, {Baum}, {Ewald}, {Hester}, {Holtzman},
  {Kristian}, {Light}, {Ligynds}, {O'Neil} Jr. \& {Westphal}}{{Lauer}
  et~al.}{1993}]{Lauer93}
{Lauer} T.~R.,  {Faber} S.~M.,  {Groth} E.~J.,  {Shaya} E.~J.,  {Campbell} B.,
  {Code} A.,  {Currie} D.~G.,  {Baum} W.~A.,  {Ewald} S.~P.,  {Hester} J.~J.,
  {Holtzman} J.~A.,  {Kristian} J.,  {Light} R.~M.,  {Ligynds} C.~R.,  {O'Neil}
  Jr. E.~J.,    {Westphal} J.~A.,  1993, \aj, 106, 1436

\bibitem[\protect\citeauthoryear{{Leigh}, {B{\"o}ker} \& {Knigge}}{{Leigh}
  et~al.}{2012}]{LGH}
{Leigh} N.,  {B{\"o}ker} T.,    {Knigge} C.,  2012, \mnras, 424, 2130

\bibitem[\protect\citeauthoryear{{Mastrobuono-Battisti}, {Perets} \&
  {Loeb}}{{Mastrobuono-Battisti} et~al.}{2014}]{mastrobuono14}
{Mastrobuono-Battisti} A.,  {Perets} H.~B.,    {Loeb} A.,  2014, \apj, 796, 40

\bibitem[\protect\citeauthoryear{{McLaughlin}, {King} \&
  {Nayakshin}}{{McLaughlin} et~al.}{2006}]{McLgh}
{McLaughlin} D.~E.,  {King} A.~R.,    {Nayakshin} S.,  2006, ApJL, 650, L37

\bibitem[\protect\citeauthoryear{{Melo} \& {Capuzzo-Dolcetta}}{{Melo} \&
  {Capuzzo-Dolcetta}}{2016}]{tostaCD16}
{Melo} I.~T.~e.,  {Capuzzo-Dolcetta} R.,  2016, Journal of Physics Conference
  Series, 689, 012008

\bibitem[\protect\citeauthoryear{{Merritt}}{{Merritt}}{2006}]{merritt06}
{Merritt} D.,  2006, \apj, 648, 976

\bibitem[\protect\citeauthoryear{{Merritt}}{{Merritt}}{2013}]{Merri13}
{Merritt} D.,  2013, {Dynamics and Evolution of Galactic Nuclei}

\bibitem[\protect\citeauthoryear{{Metzger} \& {Stone}}{{Metzger} \&
  {Stone}}{2016}]{metzger16}
{Metzger} B.~D.,  {Stone} N.~C.,  2016, \mnras, 461, 948

\bibitem[\protect\citeauthoryear{{Milosavljevi{\'c}}}{{Milosavljevi{\'c}}}{200%
4}]{Mil04}
{Milosavljevi{\'c}} M.,  2004, \apjl, 605, L13

\bibitem[\protect\citeauthoryear{{Minniti}, {Contreras Ramos}, {Zoccali},
  {Rejkuba}, {Gonzalez}, {Valenti} \& {Gran}}{{Minniti} et~al.}{2016}]{min16}
{Minniti} D.,  {Contreras Ramos} R.,  {Zoccali} M.,  {Rejkuba} M.,  {Gonzalez}
  O.~A.,  {Valenti} E.,    {Gran} F.,  2016, \apjl, 830, L14

\bibitem[\protect\citeauthoryear{{Nayakshin}, {Wilkinson} \&
  {King}}{{Nayakshin} et~al.}{2009}]{nayakshin}
{Nayakshin} S.,  {Wilkinson} M.~I.,    {King} A.,  2009, \mnras, 398, L54

\bibitem[\protect\citeauthoryear{{Neumayer} \& {Walcher}}{{Neumayer} \&
  {Walcher}}{2012}]{Neum}
{Neumayer} N.,  {Walcher} C.~J.,  2012, Advances in Astronomy, 2012

\bibitem[\protect\citeauthoryear{{Nguyen}, {Seth}, {Reines}, {den Brok}, {Sand}
  \& {McLeod}}{{Nguyen} et~al.}{2014}]{ngu14}
{Nguyen} D.~D.,  {Seth} A.~C.,  {Reines} A.~E.,  {den Brok} M.,  {Sand} D.,
  {McLeod} B.,  2014, \apj, 794, 34

\bibitem[\protect\citeauthoryear{{Perets} \& {Mastrobuono-Battisti}}{{Perets}
  \& {Mastrobuono-Battisti}}{2014}]{perets14}
{Perets} H.~B.,  {Mastrobuono-Battisti} A.,  2014, \apjl, 784, L44

\bibitem[\protect\citeauthoryear{{Petts}, {Read} \& {Gualandris}}{{Petts}
  et~al.}{2016}]{petts16}
{Petts} J.~A.,  {Read} J.~I.,    {Gualandris} A.,  2016, \mnras, 463, 858

\bibitem[\protect\citeauthoryear{{Quintero}, {Hogg}, {Blanton}, {Schlegel},
  {Eisenstein}, {Gunn}, {Brinkmann}, {Fukugita}, {Glazebrook} \&
  {Goto}}{{Quintero} et~al.}{2004}]{quintero04}
{Quintero} A.~D.,  {Hogg} D.~W.,  {Blanton} M.~R.,  {Schlegel} D.~J.,
  {Eisenstein} D.~J.,  {Gunn} J.~E.,  {Brinkmann} J.,  {Fukugita} M.,
  {Glazebrook} K.,    {Goto} T.,  2004, \apj, 602, 190

\bibitem[\protect\citeauthoryear{{Rees}}{{Rees}}{1988}]{rees88}
{Rees} M.~J.,  1988, \nat, 333, 523

\bibitem[\protect\citeauthoryear{{Reines} \& {Deller}}{{Reines} \&
  {Deller}}{2012}]{reines12}
{Reines} A.~E.,  {Deller} A.~T.,  2012, \apjl, 750, L24

\bibitem[\protect\citeauthoryear{{Rossa}, {van der Marel}, {B{\"o}ker},
  {Gerssen}, {Ho}, {Rix}, {Shields} \& {Walcher}}{{Rossa} et~al.}{2006}]{rossa}
{Rossa} J.,  {van der Marel} R.~P.,  {B{\"o}ker} T.,  {Gerssen} J.,  {Ho}
  L.~C.,  {Rix} H.-W.,  {Shields} J.~C.,    {Walcher} C.-J.,  2006, \aj, 132,
  1074

\bibitem[\protect\citeauthoryear{{Salpeter}}{{Salpeter}}{1955}]{Salpeter55}
{Salpeter} E.~E.,  1955, \apj, 121, 161

\bibitem[\protect\citeauthoryear{{Scott} \& {Graham}}{{Scott} \&
  {Graham}}{2013}]{scot}
{Scott} N.,  {Graham} A.~W.,  2013, \apj, 763, 76

\bibitem[\protect\citeauthoryear{{Seth}, {Ag{\"u}eros}, {Lee} \&
  {Basu-Zych}}{{Seth} et~al.}{2008}]{seth08}
{Seth} A.,  {Ag{\"u}eros} M.,  {Lee} D.,    {Basu-Zych} A.,  2008, \apj, 678,
  116

\bibitem[\protect\citeauthoryear{{Shankar}, {Weinberg} \&
  {Miralda-Escud{\'e}}}{{Shankar} et~al.}{2009}]{Shankar09}
{Shankar} F.,  {Weinberg} D.~H.,    {Miralda-Escud{\'e}} J.,  2009, \apj, 690,
  20

\bibitem[\protect\citeauthoryear{{Spitzer} Jr. \& {Harm}}{{Spitzer} \&
  {Harm}}{1958}]{spitzer58}
{Spitzer} Jr. L.,  {Harm} R.,  1958, \apj, 127, 544

\bibitem[\protect\citeauthoryear{{Stone}, {K{\"u}pper} \& {Ostriker}}{{Stone}
  et~al.}{2017}]{stone17}
{Stone} N.~C.,  {K{\"u}pper} A.~H.~W.,    {Ostriker} J.~P.,  2017, \mnras, 467,
  4180

\bibitem[\protect\citeauthoryear{{Stone} \& {Metzger}}{{Stone} \&
  {Metzger}}{2016}]{stone16}
{Stone} N.~C.,  {Metzger} B.~D.,  2016, \mnras, 455, 859

\bibitem[\protect\citeauthoryear{{Stone} \& {van Velzen}}{{Stone} \& {van
  Velzen}}{2016}]{stone16b}
{Stone} N.~C.,  {van Velzen} S.,  2016, \apjl, 825, L14

\bibitem[\protect\citeauthoryear{{Tremaine}}{{Tremaine}}{1995}]{Tremaine95}
{Tremaine} S.,  1995, \aj, 110, 628

\bibitem[\protect\citeauthoryear{{Tremaine}}{{Tremaine}}{1976}]{Trem76}
{Tremaine} S.~D.,  1976, \apj, 203, 345

\bibitem[\protect\citeauthoryear{{Turner}, {C{\^o}t{\'e}}, {Ferrarese},
  {Jord{\'a}n}, {Blakeslee}, {Mei}, {Peng} \& {West}}{{Turner}
  et~al.}{2012}]{Turetal12}
{Turner} M.~L.,  {C{\^o}t{\'e}} P.,  {Ferrarese} L.,  {Jord{\'a}n} A.,
  {Blakeslee} J.~P.,  {Mei} S.,  {Peng} E.~W.,    {West} M.~J.,  2012, \apjs,
  203, 5

\bibitem[\protect\citeauthoryear{{Urry} \& {Padovani}}{{Urry} \&
  {Padovani}}{1995}]{UrPa}
{Urry} C.~M.,  {Padovani} P.,  1995, \pasp, 107, 803

\bibitem[\protect\citeauthoryear{{van den Bosch}, {Gebhardt}, {G{\"u}ltekin},
  {van de Ven}, {van der Wel} \& {Walsh}}{{van den Bosch}
  et~al.}{2012}]{VdeBo12}
{van den Bosch} R.~C.~E.,  {Gebhardt} K.,  {G{\"u}ltekin} K.,  {van de Ven} G.,
   {van der Wel} A.,    {Walsh} J.~L.,  2012, \nat, 491, 729

\bibitem[\protect\citeauthoryear{{Vink{\'o}}, {Yuan}, {Quimby}, {Wheeler},
  {Ramirez-Ruiz}, {Guillochon}, {Chatzopoulos}, {Marion} \&
  {Akerlof}}{{Vink{\'o}} et~al.}{2015}]{vinko15}
{Vink{\'o}} J.,  {Yuan} F.,  {Quimby} R.~M.,  {Wheeler} J.~C.,  {Ramirez-Ruiz}
  E.,  {Guillochon} J.,  {Chatzopoulos} E.,  {Marion} G.~H.,    {Akerlof} C.,
  2015, \apj, 798, 12

\bibitem[\protect\citeauthoryear{{Walcher}, {B{\"o}ker}, {Charlot}, {Ho},
  {Rix}, {Rossa}, {Shields} \& {van der Marel}}{{Walcher}
  et~al.}{2006}]{walcher06}
{Walcher} C.~J.,  {B{\"o}ker} T.,  {Charlot} S.,  {Ho} L.~C.,  {Rix} H.-W.,
  {Rossa} J.,  {Shields} J.~C.,    {van der Marel} R.~P.,  2006, \apj, 649, 692

\bibitem[\protect\citeauthoryear{{Wang} \& {Merritt}}{{Wang} \&
  {Merritt}}{2004}]{wang04}
{Wang} J.,  {Merritt} D.,  2004, \apj, 600, 149

\bibitem[\protect\citeauthoryear{{Yang}, {Paragi}, {van der Horst}, {Gurvits},
  {Campbell}, {Giannios}, {An} \& {Komossa}}{{Yang} et~al.}{2016}]{yang16}
{Yang} J.,  {Paragi} Z.,  {van der Horst} A.~J.,  {Gurvits} L.~I.,  {Campbell}
  R.~M.,  {Giannios} D.,  {An} T.,    {Komossa} S.,  2016, \mnras, 462, L66

\bibitem[\protect\citeauthoryear{{Yang}, {Zabludoff}, {Zaritsky}, {Lauer} \&
  {Mihos}}{{Yang} et~al.}{2004}]{yang04}
{Yang} Y.,  {Zabludoff} A.~I.,  {Zaritsky} D.,  {Lauer} T.~R.,    {Mihos}
  J.~C.,  2004, \apj, 607, 258

\bibitem[\protect\citeauthoryear{{Zabludoff}, {Zaritsky}, {Lin}, {Tucker},
  {Hashimoto}, {Shectman}, {Oemler} \& {Kirshner}}{{Zabludoff}
  et~al.}{1996}]{zabludoff96}
{Zabludoff} A.~I.,  {Zaritsky} D.,  {Lin} H.,  {Tucker} D.,  {Hashimoto} Y.,
  {Shectman} S.~A.,  {Oemler} A.,    {Kirshner} R.~P.,  1996, \apj, 466, 104

\end{thebibliography}
}

\end{document}